# The electricity system value of the local acceptance of onshore wind in Europe

James Price[1], Guillermo Valenzuela-Venegas[2], Oskar Vågerö[2], Marianne Zeyringer[2], Monika Bucha[3], Ruihong Chen[4], Adrienne Etard[5], Andrea N. Hahmann[6], Alena Lohrmann[4,7], Russell McKenna[4,8], Christian Mikovits[9], Evangelos Panos[4,8], Meixi Zhang[8], Luis Ramirez Camargo[10]

[1] UCL Energy Institute, University College London, London, UK
[2] Department of Technology Systems, University of Oslo, Oslo, Norway
[3] Kelso Institute Europe, Berlin, Germany
[4] Chair of Energy System Analysis, ETH Zurich, Zurich, Switzerland
[5] Biodiversity, Ecology, and Conservation Research Group, International Institute for Applied Systems Analysis, Laxenburg, Austria
[6] Department of Wind and Energy Systems, Technical University of Denmark, Roskilde, Denmark
[7] Iceland School of Energy, Reykjavik University, Reykjavik, Iceland
[8] Laboratory for Energy Systems Analysis, PSI Centers for Nuclear Engineering and Sciences & Energy and Environmental Sciences, Paul Scherrer Institute, Villigen, Switzerland
[9] Institute of Sustainable Economic Development, University of Natural Resources and Life Sciences, Vienna, Austria
[10] Copernicus Institute of Sustainable Development, Utrecht University, Utrecht, the Netherlands

**Abstract**

The large-scale deployment of wind power is central to Europe's energy transition but faces challenges due to its social and environmental impacts on communities. Here we assess how the tolerance of local stakeholders to such impacts translates across spatial scales to shape the cost and design of the continent's net-zero electricity system using a soft-linked modelling framework. We find that lower impact tolerance can reduce the role of onshore wind in Europe reaching net-zero by up to 84% relative to a future where wind enjoys higher acceptance, with other low carbon sources needing to be scaled up to compensate. This translates into total European electricity system costs increasing by between 2-14% while some countries see costs escalating by 20% or more. Our results show that the local acceptance of onshore wind is a key structural driver of the system and highlight the system value of policies to promote it.

**Keywords:** Wind energy, Social and environmental impacts, Energy system transition, Energy systems modelling

# Main

The rapid decarbonisation of global energy systems needed to limit climate change to well below 2°C will rely heavily on key renewable sources like wind power. However, despite its economic and technical viability, the large-scale deployment of wind faces considerable obstacles, many of which manifest at the local level and drive resistance to the construction of new infrastructure. Such challenges can then feedback to the national and even global context by impeding the delivery of the energy transition and its net-zero objectives. Here we explore this interplay in the European context using a range of important local concerns pertaining to wind that we categorise into two dimensions, social and environmental.

On the social side, challenges include the visual impact associated with new infrastructure[1–3], which can be particularly exacerbated in more scenic areas[4]. Other annoyance related aspects, such as the noise turbines produce while operating[1,5,6] and the shadow flicker caused by blade rotation[7–9], can also undermine the acceptance of wind power. Greater local sensitivity to these impacts reduces the number of acceptable sites available for wind.

Environmental conservation objectives can also put pressure on the land available for new wind infrastructure. Common threads include avoiding protected conservation areas[10–12] and the need to minimise the impact on flying species (e.g. birds, bats) resulting from collision with turbines[13], especially if those species are vulnerable due to their ecological characteristics (e.g. generation length) and conservation status. Ecological concerns also extend to preserving important habitats like forests[14–16] and peatlands[17,18], both for biodiversity and climate goals[19,20].

Studies have integrated such social and environmental impacts to quantify the trade-offs between different wind siting priorities, e.g. environmental protection, under predefined wind deployment targets[15,16,21–23], but have neglected their implications for wider system planning. National or sub-national energy systems models, which are crucial tools to underpin energy and climate policymaking, have also incorporated such factors to capture the interaction between local siting and the overall design and costs of future systems[24–32]. These works include a patchwork of the social and environmental considerations detailed above.

However, to date, no study has embedded these factors in an energy system model for Europe to understand how they could influence the continent's net-zero goals. Indeed, continental and global systems models often simplify or fail to capture key factors entirely, while national analyses that do account for them do not allow for a consistent assessment of trade-offs between countries. Moreover, the important aspects of landscape aesthetics and the vulnerability of flying species have not been combined with more typical siting restrictions (e.g. setbacks from urban areas) in a multi-country energy system planning framework. Furthermore, the implications of sustaining existing wind planning practices for the continent's net-zero electricity system have not been assessed. Thus, it is critical to close these gaps and quantify how the local social and environmental impacts of wind translate across spatial scales to influence electricity system design at the national and European level.

Here, we address this by investigating how the tolerance of the local impacts of wind power can shape the overall design of Europe's future electricity system. We develop and apply a modelling framework that captures the long-term transition of the whole European energy system and spatially detailed electricity system planning. Wind power deployment in this framework is constrained by land availability scenarios, which represent potential futures in which stakeholders across the continent have different levels of sensitivity to its social and environmental impacts. Lower tolerance of these impacts results in less land available for wind siting. Compared to a future where acceptance is high, low impact tolerance reduces the capacity and generation potential of onshore wind by 95% and drives its re-location more towards the periphery of Europe in areas such as Iberia and the Nordics. Total annual electricity system costs for the continent increase by between 2-14% relative to the most tolerant scenario, with a number of countries seeing a 20% or more increase in costs. Our results highlight the system value of the local acceptance of onshore wind and therefore policies that seek to promote it in communities across Europe.

## Modelling wind power futures for Europe

The social acceptance of wind power may be understood based on its socio-political, market and community acceptance according to the framework of Wüstenhagen[33]. With both socio-political and market acceptance high across Europe[34,35] a key determinant of a project's acceptability will be its local impacts. Therefore, in this study we conceptualise acceptance as equivalent to community acceptance, i.e. the specific approval of wind projects by local stakeholders. As such, we assume that both the cumulative amount and location of land available for the siting of onshore wind in Europe depends on the tolerance of communities to the impacts, both real and perceived, it has on people and ecosystems locally. Taking the visual impact of wind farms as an example, a lower tolerance would imply that projects would need to be sited further from urban areas in order to be acceptable while a higher tolerance would mean the opposite. While we recognise the complexity of social acceptance and that the three acceptance dimensions are closely intertwined, focusing on this definition of acceptance facilitates our exploratory geospatial analysis.

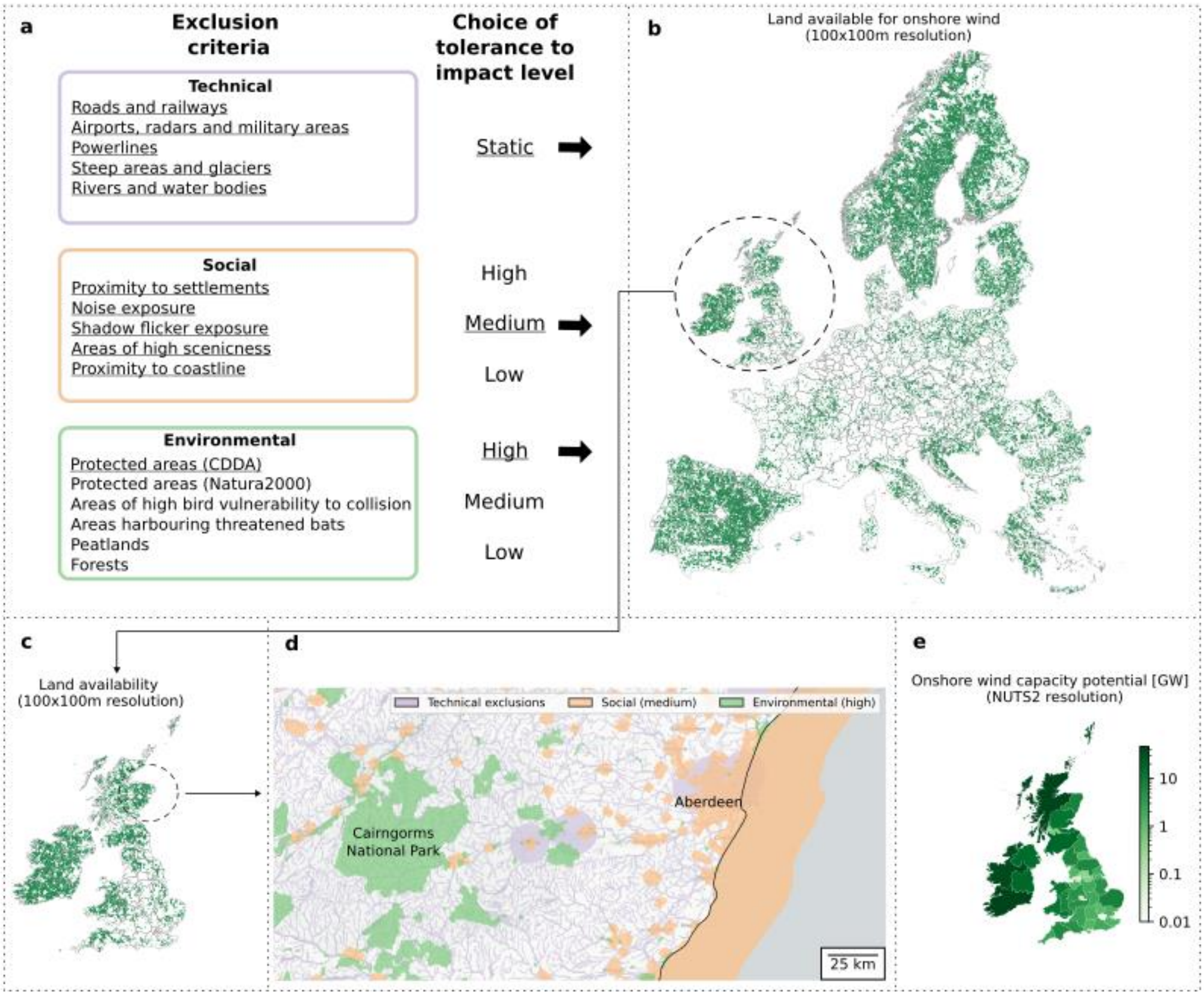

**Figure 1. Overview of the land availability modelling framework.** Panel a) shows the exclusion criteria, with the criteria included in the medium-high (where tolerance to social impacts is medium and environmental is high) scenario underlined and displayed in all the subsequent panels. CDDA is the Common Database on Designated Areas for officially designated protected areas. Similarly, Natura 2000 is a network of protected areas in the European Union (see Methods for further details). Panel b) shows the land available for onshore wind deployment (in green) for this scenario for all of Europe, with a 100 m x100 m resolution. To illustrate the detail of the high-resolution land availability scenarios and data, we zoom in on the United Kingdom and Ireland (panel c) and Aberdeen city and Aberdeenshire (panel d) to showcase the three categories of exclusion criteria. Finally, the resulting onshore wind capacity potential in GW at NUTS2 resolution is shown in panel e) with a logarithmic scale.

We develop three levels of tolerance across both our social and environmental impact dimensions from a more permissive case (high) through a continuation of planning outcomes aligned with current national planning practices (medium) to a more restrictive state (low). Each impact dimension is further broken down into a collection of relevant categories, e.g. setback distances to settlements in the social dimension, which are gradated in accordance with tolerance level. The quantification of each category across our scenario levels is based on current national planning regulations, informed by wind deployment patterns to date or existing literature. Therefore, our scenarios are an evidence-based assessment of the range of possible outcomes that the planning system in each country we model may deliver in future. This results in nine European-wide high resolution land availability scenarios, all of which also incorporate important technical constraints such as the maximum slope of sites suitable for wind farms. Our scenarios focus predominately on onshore wind given its well documented siting challenges but do also include social, environmental and technical aspects relevant to offshore wind. See Fig. 1 for an example of the land availability framework. The scenario naming convention is firstly tolerance to social impacts and then to environmental impacts, e.g. medium-high.

To explore the implications of these scenarios for the future of wind power in Europe, we couple the long-term whole energy system model JRC-EU-TIMES with the spatially detailed electricity system model highRES-Europe. The former model optimises the structural evolution of the energy system to achieve Europe's net-zero climate policy objectives, including the extent to which the system electrifies and the decarbonisation of electricity generation in each country. These boundary conditions for 2050 feed our electricity system model, which designs spatially detailed net-zero aligned electricity systems for 25 EU countries plus the UK, Switzerland and Norway. We use this framework to elaborate the system cost and design trade-offs across these scenarios and quantify the contribution of wind energy to Europe's net-zero future. For further details, see the Methods.

## Tolerance of social and environmental impacts profoundly shapes Europe's wind potential

Communities' tolerance of the social and environmental impacts of wind power strongly controls the availability of land to site onshore wind in our scenarios with a ~96% reduction in European-wide onshore wind capacity (~5000 GW to ~200 GW, Fig. 2) when moving from the highest (high-high) to lowest tolerance case (low-low). If planning process outcomes to 2050 reflect recent experience (medium-medium) then ~50% of the capacity of the high-high scenario are available.

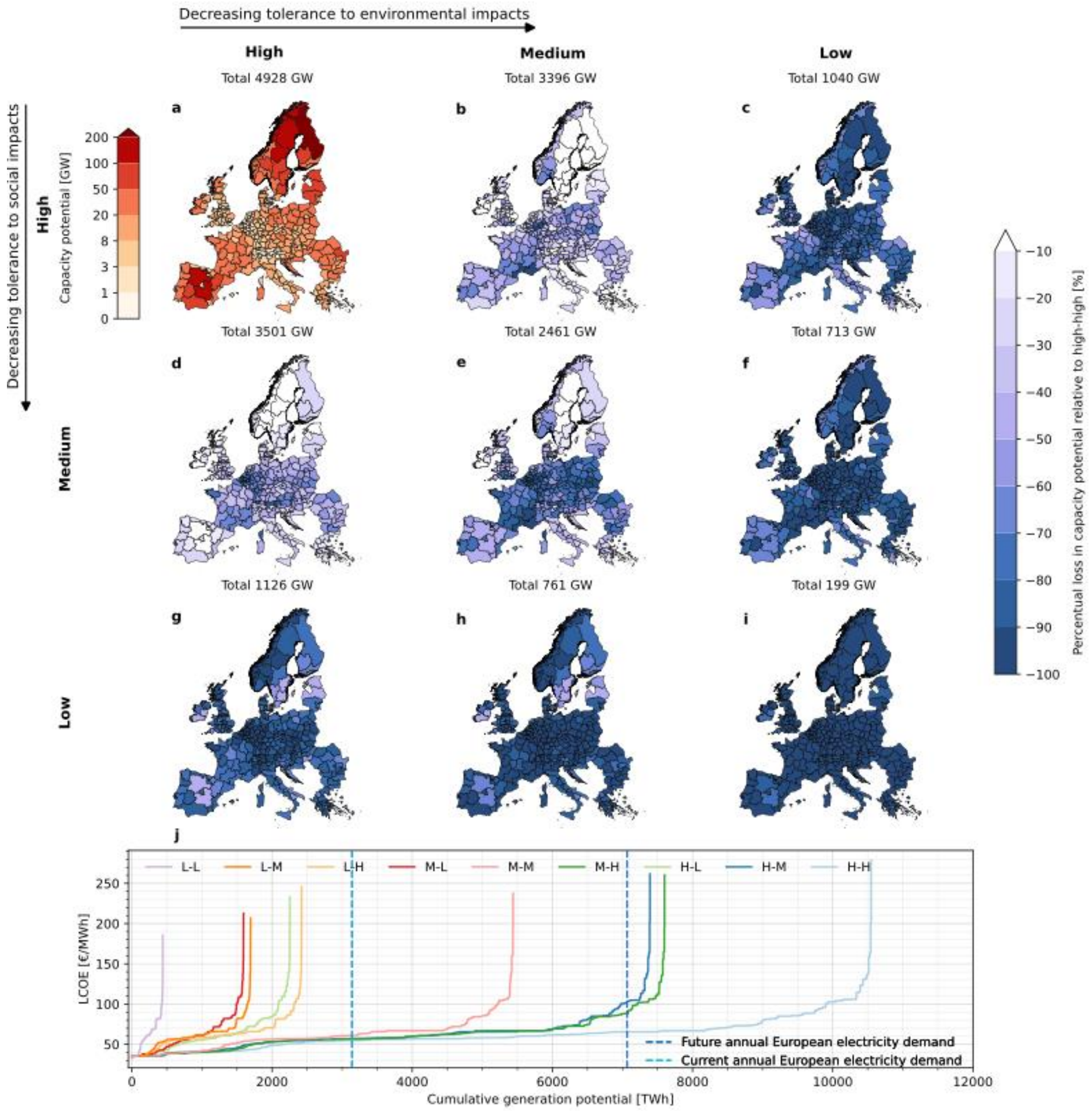

**Figure 2. New capacity potential for wind energy deployment across nine scenarios of tolerance to social and environmental impacts and associated supply curves.** Panel a) Shows the absolute new capacity potential (in GW) in the most permissive scenario (high-high). Panels b-i) show the percentual capacity potential loss compared to high-high, illustrating the regional differences at NUTS2 resolution as land availability progressively reduces. On top of every panel is the total European capacity potential (in GW). Panel j) shows supply curves for the nine scenarios, illustrating not only the absolute capacity reduction of the scenarios, but also the wind quality of removed areas. The two vertical lines represent current[36] and modelled future European electricity demand. Abbreviations should be read as level of social dimension followed by level of environmental dimension (i.e. L-M is low social and medium environmental). The technical exclusions (Methods - Technical exclusions) are active across all scenarios. High resolution versions of the figure, for ten European regions, are available in Supplementary Information Note 1 (Supplementary Figure S1.1 - S1.10) along with capacity potential at the country-level (Supplementary Table S1.1) and NUTS2-level (Supplementary Data S1).

Regions such as Iberia, parts of the Nordics, the UK and Ireland tend to retain a greater number of siting opportunities for wind in less permissive (low tolerance) cases with more centrally located countries, e.g. France, Germany and Belgium, seeing large reductions in available land area. This is driven by a combination of social (e.g. settlement density) and environmental (e.g. protected areas) factors. At the continental level, a lower tolerance of environmental impacts reduces capacity potential more than a comparable level of social impact tolerance, although the effect is modest.

The onshore wind supply curves for Europe across our scenarios highlight the sizable reduction in cumulatively available wind energy moving from high-high to low-low, i.e. ~11600 TWh/yr to ~500 TWh/yr (Fig. 2j). Again, this indicates the dramatic effect that the sensitivity of local communities to its impacts can have on the amount of wind energy available. Furthermore, this panel demonstrates that high quality, i.e. windy sites, are preferentially lost as stakeholder impact tolerance reduces as indicated by the steepening gradient of the curves. This implies that the best sites for wind power are often located in socially or environmentally sensitive areas. As with the capacity potential, less tolerance of environmental impacts generally leads to marginally less total available wind energy than social impacts at the equivalent level.

Next we assess how site availability shapes onshore wind's spatial distribution and cumulative deployment in net-zero aligned European electricity systems in 2050. From Fig. 3 we see a range of between 1446 (high-high) and 238 GW (low-low), an 84% drop in installed capacity across our scenarios. Spatially, the least cost solution moves from large concentrations of wind power in certain high quality NUTS2 regions, particularly in Northern, North-western and Western Europe, e.g. in France and the UK, to a more evenly distributed picture across these areas. Large concentrations of capacity tend to follow the trends observed for land availability and shift to (or are maintained in) Iberia, parts of the Nordics and Ireland and away from major demand centres like France and Germany. This spatial reallocation of capacity reflects the optimal system level response to the tightening restrictions on wind throughout Europe, with areas previously seen as not cost-effective becoming utilised.

In general, scenarios which represent substantially lower tolerance of environmental or social impacts compared to the other dimension (Fig. 3c,g) show broadly similar optimal spatial deployment patterns. However, areas in Southern Spain and Northern Finland have notably less wind capacity when the tolerance of environmental impacts is reduced. For the former this is likely to be the case because this scenario limits wind deployment in areas with bird species that are particularly vulnerable to collisions with turbines, e.g. the Eurasian griffon vulture[37,38].

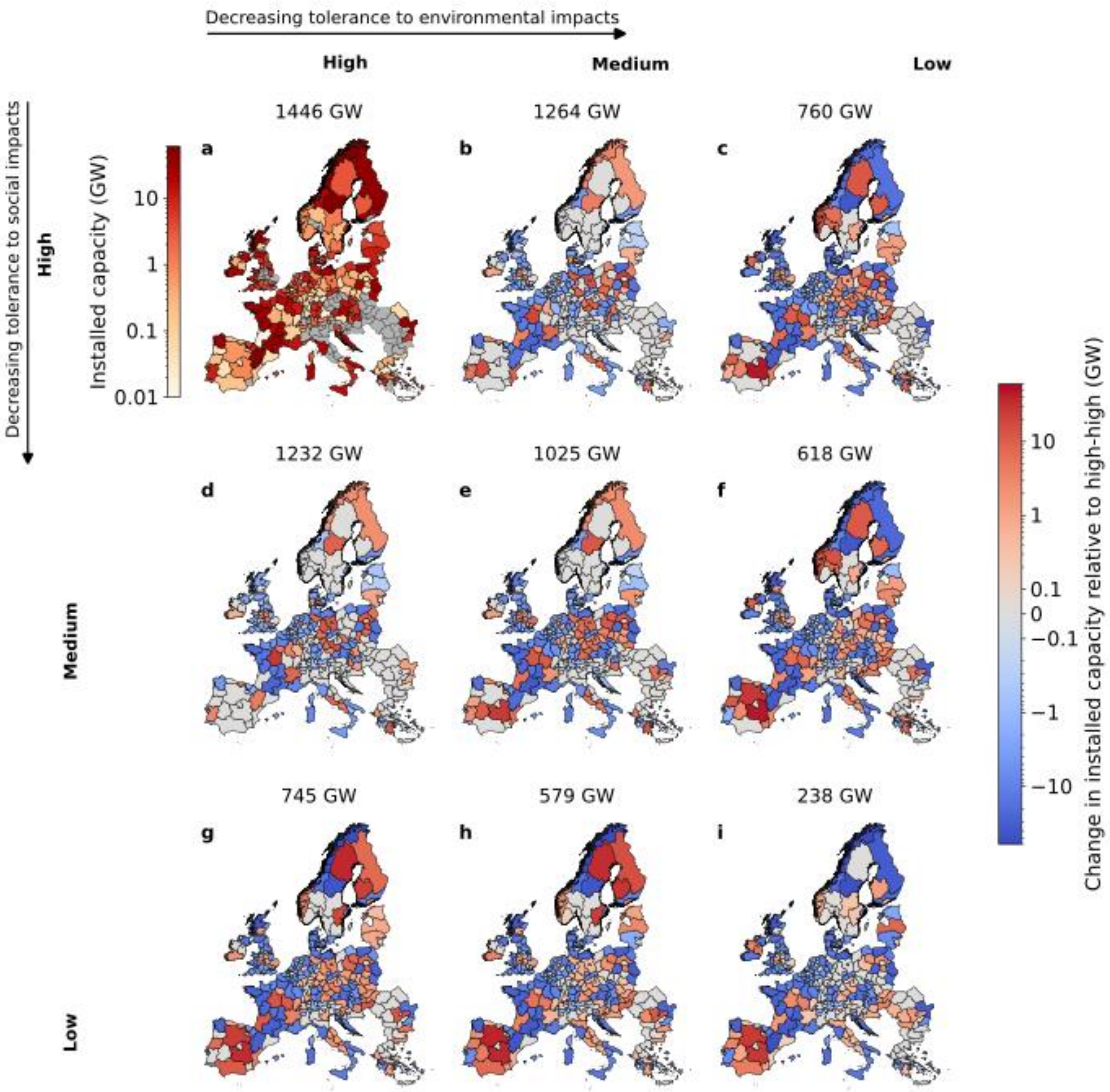

**Figure 3. Spatial (NUTS2) distribution of onshore wind capacities in 2050.** Panel a) shows the installed capacity (in GW) for the high social and high environmental impact tolerance case on a logarithmic scale. The remaining panels show the change in installed capacity relative to this scenario. Total installed capacity across the EU25+3 countries modelled here is shown in the title of each panel. For details on country-level results and how these compare to present European onshore wind energy deployment, see Supplementary Figure S1.11.

## System design and costs strongly influenced by availability of onshore wind

A diminished role for onshore wind in mid-century European electricity systems has substantial implications for the rest of the system, with larger contributions required from other low carbon options like offshore wind, solar PV and, in the least permissive cases, nuclear (Fig. 4b). In low-low, more than 1200 GW of onshore wind is lost and replaced by ~450 GW of offshore wind and ~500 GW of solar PV. It is worth noting that these technologies themselves may experience obstacles at the local level which could impede their deployment and compensatory role for losses in onshore wind.

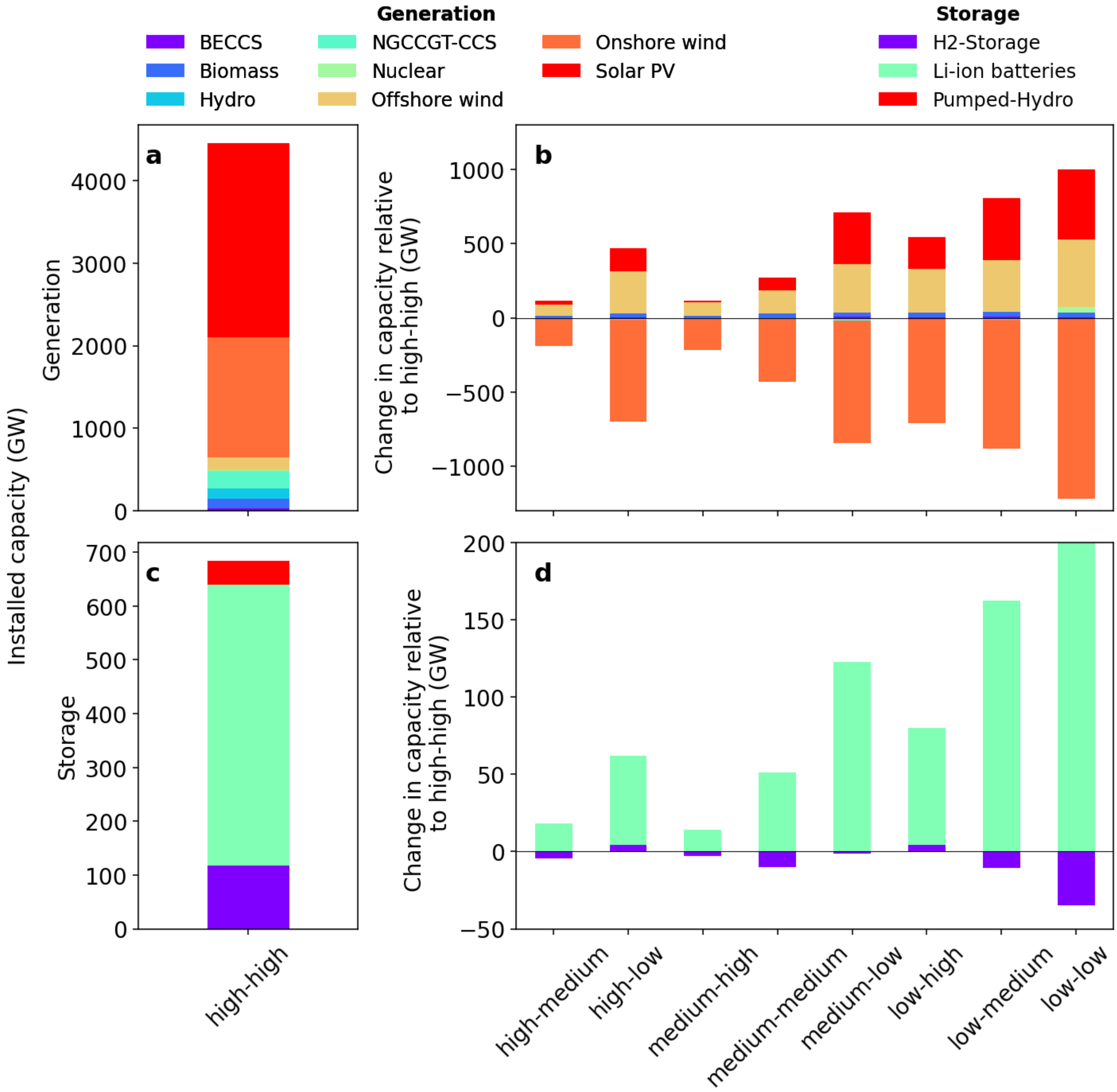

**Figure 4. Electricity system design and change between scenarios in 2050.** Cumulative installed capacity of generation (a) and storage (c) in the EU25+3 countries modelled here for the scenario with high social and high environmental impact tolerance (high-high). Panels b) and d) show how these capacities change across the social and environmental impact tolerance scenarios. BECCS = Bioenergy with Carbon Capture and Storage, NGCCGT-CCS = Natural gas Combined-Cycle Gas Turbine with Carbon Capture and Storage, H2-Storage = Hydrogen storage in metal tanks with $H_2$ created by electrolysis and converted back to power using either an open-cycle or combined-cycle gas turbine.

Fig. 4c,d demonstrates that a net-zero European electricity system relies heavily (in power generation terms) on battery storage, supported by longer duration assets, and this reliance only grows as the role of onshore wind declines across our scenarios. This is partially

explained by the synergy between batteries and solar PV. Additionally, hydrogen stored in tanks with a discharge duration of ~45 hours has a modest but consistent role to play.

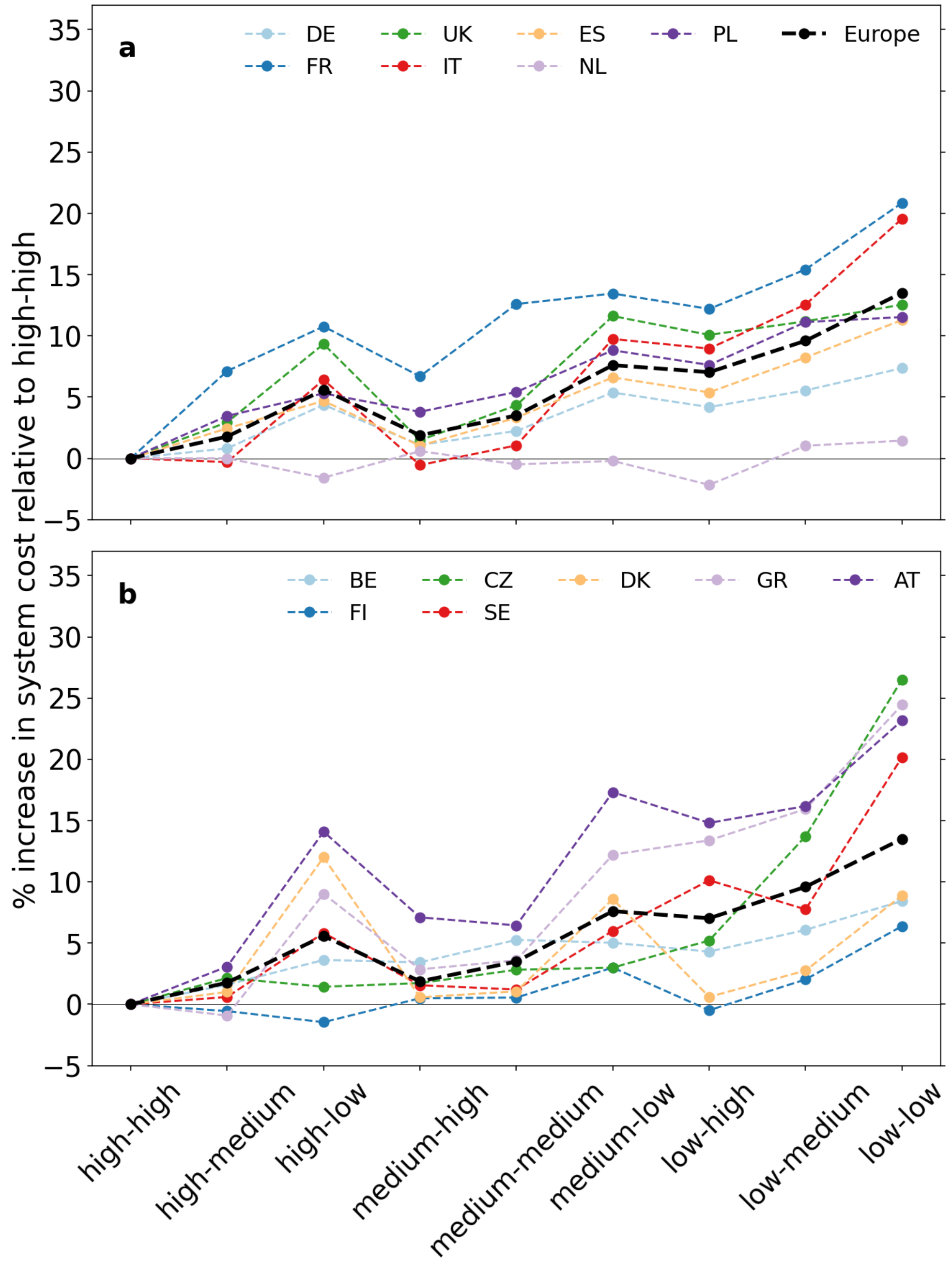

**Figure 5. Change in total annual systems costs for Europe and selected countries in 2050.** Panels a) and b) show the relative total annual system cost increase for the top 14 most expensive national systems in our modelling, cumulatively amounting to 90% of Europe's total system cost. On both panels cost increases for the European electricity system as a whole are also plotted in black. The total system cost for Europe includes annualised capex for generation, storage and transmission together with fixed operating and maintenance and variable costs (start-up, fuel and operating and maintenance). Country level total annual system costs are derived using equation 3 from Ref.[39].

These substantial changes in the design of Europe's electricity system in 2050, driven by the progressive reduction in the amount and quality of sites for onshore wind deployment across our scenarios, result in increases in total annual system costs (Fig. 5). For Europe as a whole, these increases are relatively modest for all combinations of high and medium, with medium-medium being ~3.5% or 24 €$_{2024}$ bn/yr more expensive than the most optimistic case high-high. However, bringing the low tolerance level of either dimension into play leads to ~6% or more higher costs as the available space for onshore wind is further restricted. In the most pessimistic case modelled here, this trend ultimately reaches a ~14% increase in total system costs. This means that low tolerance to the social and environmental impacts of wind across Europe, which substantially diminishes its role in delivering net-zero, can lead to an extra ~91 €$_{2024}$ bn/yr in expenditure on the continent's electricity system.

A sensitivity analysis was conducted to understand how various core modelling assumptions may affect the escalation of total system costs (see Supplementary Note 2). Increases in total costs are 3-4% and 10-15% for medium-medium and low-low, respectively, across these sensitivity cases. Greater amounts of $CO_2$ sequestration capacity, higher solar PV deployment and the addition of floating offshore wind are found to moderate the impact of onshore wind being limited. However, even in these cases, which come with their own feasibility challenges, onshore wind still has substantial value to the system.

Total cost increases at the European level are not distributed evenly between countries (Fig. 5a,b). Indeed, the Netherlands see system costs decrease in some scenarios and the expenditure in Germany, by far the costliest electricity system in our modelling, only grows by at most ~7%. However, country system costs typically increase markedly more than for Europe as a whole in, for example, France and Italy, with the latter particularly sensitive to onshore wind expansion being more constrained than in a scenario that aligns with today's siting patterns (medium-medium). System cost growth for the UK tracks at or sometimes well above that of Europe combined, as would be expected given the importance of wind power to that nation's decarbonisation, while costs in Czechia and Austria increase by nearly 30% and 25%, respectively. Overall, it is clear that for some countries onshore wind plays a crucial role and its deployment being constrained by the lower tolerance of its impacts has significant implications for delivering a cost-effective net-zero electricity system in 2050.

# Discussion

We have demonstrated that the tolerance of local communities across Europe to the social and environmental impacts of wind power has substantial implications for its role in the continent's net-zero electricity system. Clear trade-offs are apparent with lower acceptance, operationalised by greater impact mitigation such as larger setback distances from settlements, driving higher costs and a smaller role for onshore wind. This highlights how local acceptance must be accounted for when modelling energy futures to avoid misleading results which may underestimate system costs and even fail to deliver net-zero in the real world. It also indicates the system wide value of policies which can help mitigate the impacts of onshore wind and improve its local acceptance, thereby supporting its integration into landscapes. As we have found, not doing so results in important consequences that must be considered.

A more constrained future role for onshore wind implies that other low carbon sources such as offshore wind, solar PV and nuclear will need to be scaled appropriately to compensate and meet future electricity demand. However, these options also have with their own impacts which can affect their acceptance, the majority of which we have not considered. Ground mounted solar PV, for example, can suffer from some of the same local issues as onshore wind, such as its perceived visual impact on the landscape[27,40], while nuclear has a host of attendant social and environmental concerns[41,42]. Therefore, the system cost ramifications we report may underestimate the implications of the low acceptance of onshore wind for Europe and in future the assessment should be expanded to also include the impacts of other low carbon options.

Furthermore, the higher system costs we find when local community acceptance is low, particularly in countries with a strong reliance on wind, would mean the energy transition places a greater economic burden on society. This in turn would likely translate into negative implications for the socio-political feasibility of the transition in general[43,44]. In the UK, for example, the consensus around net-zero is under pressure due to its costs being politically contested, even if the estimated net cost of the nation's transition has reduced by 73% in just five years[45].

However, it is important to recognise that local acceptance is not static and is subject to change given appropriate policy making. Recently, a number of European countries have begun to introduce financial participation and compensation schemes requiring wind developers to compensate local communities through fixed payments or by offering opportunities for financial participation[46]. Numerous studies[47–51] demonstrate the positive effects of such schemes on local acceptance, with several[52–55] showing that residents seem to prefer community compensation over individual payments. Evidence also indicates that citizen involvement in how municipal funds are distributed is key[48,53,56], e.g. school renovation or measures to protect birds. Other successful interventions to improve acceptance include participation in project planning and siting[47,57,58]. Depending on their design, such schemes can thus weave in strong elements of participatory and distributive justice by granting citizens agency and control. Mitigating the impacts on wildlife via options such as careful site selection, blade painting and on-demand shutdown to avoid collisions[59,60], which has been shown to only minimally decrease total electricity output, can also crucially de-conflict projects and promote acceptance. Therefore, policies and approaches to foster the community acceptance of wind power may support the feasibility of the energy transition by enhancing its inclusivity and empowering local stakeholders, together with reducing costs as we have shown.

We recognise our study has caveats. While our scenarios are based on existing regulations where possible, the literature and our own modelling, they do not engage directly with local communities in a participatory manner and generally do not capture socio-cultural differences around impact tolerance within and between countries. This is a fundamental limitation of continental-scale analyses, however our study can inform and guide participatory processes. We primarily focus on only one aspect of social acceptance, that of the local community, and do not represent the wider socio-political or market acceptance of wind, although evidence shows both of these dimensions see high acceptance in Europe[35,61]. Our approach to operationalise acceptance is deterministic and going forward it would be of interest to move to a more probabilistic framing. Electricity demand is inelastic and inflexible in our electricity system modelling and, while there is significant uncertainty as to the uptake of demand side measures, it would be valuable to relax this assumption in future work.

To conclude, our results demonstrate that community acceptance of onshore wind is not just a local planning concern, but a structural determinant of Europe’s net-zero electricity system. Policies that promote the acceptance of wind act at a system-level to shape its design and help avoid locking Europe into higher-cost energy futures. Therefore, it is critical for policymakers to recognise that supporting community acceptance fosters a more just and feasible energy transition that is also more cost-effective.

# Methods

### Scenario narratives

Our scenarios span three levels of local tolerance (low, medium and high) across both social and environmental impact dimensions and, as such, represent a broad range of possible futures for wind deployment in Europe. A high tolerance to both its social and environmental impacts (high-high) reflect the most permissive scenario and a low tolerance (low-low) reflects the most restrictive scenario.

*The high tolerance scenarios* are anchored in a future where onshore wind energy is well tolerated by citizens, and community acceptance is high. This necessitates ameliorating negative impacts and bringing associated benefits to local communities and not only negative externalities. There are several examples of countries and periods where onshore wind deployment has happened without much conflict. For instance, Denmark has, in the past, been considered an international frontrunner in wind power development since the 1990s[62], with the foundation laid by small-scale wind farm cooperatives in the 1970s and 1980s[63].

*The medium tolerance scenarios* are intended to approximate the accumulation of planning practices and policies to date and to represent their implications for future onshore wind deployment. Both formal regulations/guidelines and informal planning practices vary considerably between countries and introduce spatial heterogeneity. While this is subject to change in the future, the medium scenarios make use of national regulatory data as well as inferred spatial deployment patterns to extrapolate a future where wind energy deployment follows the aggregation of past practices (see the following sections for further details).

*The low tolerance scenarios* are the most restrictive and pessimistic scenarios for onshore wind deployment and represent a future where acceptance of the technology's impacts is low. In this future, European onshore wind potential is severely constrained. This perspective is grounded in situations where onshore wind impacts are perceived as high, and the technology is heavily contested. Such situations have a number of historical precedents in Europe. We include evidence of such situations for Norway, Austria, the United Kingdom and Denmark in Supplementary Note 3. Drawing on these examples, the low tolerance scenario is intended to recognise and elaborate a future where the deployment of wind power becomes highly contested across the continent.

### Scenario implementation

These scenario narratives, and the futures that they illustrate, are implemented in two dimensions, *social* and *environmental*, with three levels each resulting in a total of nine scenarios. Data used for both the social and environmental dimension is a combination of already publicly available datasets and newly developed European-wide spatially granular datasets from the WIMBY project (https://wimby.eu/). These new datasets include an assessment of landscape scenicness[64], bird vulnerability to collisions with wind turbines[65] and pseudo-species richness for threatened bat species[66], regulatory setback distances from settlements[67] as well as shadow flicker[68] from onshore wind turbines. Additionally, wind farm footprint data, derived from a novel satellite imagery (Copernicus Sentinel-2) approach[69], developed to identify land- and sea-use change, were used to inform the assessment of aggregate wind energy planning decisions across countries.

Below we provide a summary of the data and implementation used for our social and environmental impact dimensions. For a full breakdown of the quantitative implementation of each level see Supplementary Table S3.1 and Supplementary Table S5.7 for the detailed spatial differentiation in the medium level scenario.

*Environmental dimension*

The environmental dimension includes protected areas from the Common Database on Designated Areas[70,71] (CDDA, which combines data from 2023 for the EU and 2019 for the UK) and Natura 2000[72] datasets, peatland and forest data from CORINE[73], threatened bat pseudo-species richness (10 species from Ref.[74] spatially modelled in Ref.[66]), and bird vulnerability to collisions[65]. The latter novel dataset represents the number of species vulnerable to collisions at a 10km x 10km resolution throughout Europe. Vulnerability is assessed for 108 bird species known to fatality collide with wind power infrastructure and is defined based on key ecological characteristics (generation length, clutch size and area of habitat suitability). Our study combines pseudo-species richness maps of "High risk" and "High latent risk" species from Ref.[65].

In the case where communities have a high tolerance to environmental impacts, only the protected areas from the CDDA are included, and the other exclusions are not applied at this level. In the CDDA dataset, protected areas are categorised based on the International Union for the Conservation of Nature (IUCN) management categories Ia to VI, and at this level we exclude Ia (strict nature reserve), Ib (wilderness area), II (national park), III (natural monument or feature) and IV (habitat or species management area) from wind development. We assume that these areas are largely protected from human activities, despite actual important variation in management intensity and human activity across protection categories[75,76].

The medium level of tolerance to environmental impacts approximate today's planning practices and therefore contain country-specific exclusions generated by an analysis of the overlap of existing wind farm footprints (based on Ref.[69]) and the various environmental datasets mentioned above. Essentially, a category is excluded from wind energy deployment in a country if less than 15% of existing wind farms overlap with the category, as it indicates that although there is no formal legislation preventing deployment, it has been avoided previously. The detailed results of the analysis are shown in Supplementary Note 5, Table S5.2-S5.4 and Table S5.6.

A low level of tolerance to environmental impacts assumes all CDDA protected areas and Natura 2000 areas are excluded in all countries, including a 2km buffer distance around these sites based on Ref.[77]. Areas with the 40% highest vulnerable bird and bat pseudo-species richness in each country are excluded to minimise the impact on these species. Furthermore, no wind development is permitted on peatlands and in forests.

*Social dimension*

The social dimension includes setback distances from settlements based on existing regulations[67] as well as noise[78,79] and shadow flicker[68] impacts from wind turbines. The regulatory dataset, collected through semi-structured interviews with wind energy experts, presents a comprehensive overview of the regulations and guidelines for all EU25+3 countries concerning setback distances that must be complied with when constructing new onshore wind projects. Data on shadow flicker impacts are from the open-source WIMBY_SF tool[68], which is used to estimate the setback distance required to ensure that shadow flicker exposure is below a certain threshold. Lastly, data on landscape scenicness[64] are obtained using a machine learning model (Extreme Gradient Boosting classifier) trained on scenicness data (https://scenicornot.datasciencelab.co.uk/) from Great Britain. The model utilises various features, including naturalness, ruggedness, remoteness and human impact to classify landscapes at a 1x1km$^2$ resolution as "other", "scenic", and "highly scenic".

For the high level of tolerance to social impacts, the regulatory setback distances are set to 200m, which is the lowest distance identified from the regulatory dataset[67]. Similarly, the setback distances for noise and shadow flicker are set to 250m and 84m, respectively. These distances are defined to keep a median noise level below 55 dB(A)[79], to prevent detrimental health effects according to the EU Environmental Noise Directive[80], and to ensure an exposure time to shadow flicker of approximately 50 hours per year, considering one turbine hub height[68]. The buffer distance from coastlines is set to 6 nautical miles (11,112m), in line with the less restrictive scenario from Ref.[25]. No restrictions based on scenicness are applied at the high level.

The medium level of tolerance to social impacts introduces country-specific exclusions, informed by the regulatory dataset[67] and the overlap analysis between existing wind farm footprints and the dataset on landscape visual impact. Country-level regulatory data is applied for the setback distances from settlements, whereas the noise setback distance is set to 500m, corresponding to the typical value in Ref.[78] and to keep noise below a maximum of 45-50 dB(A). Shadow flicker setback distance is set to 1250m, corresponding to the distance required to limit the shadow flicker exposure to 30 hours per year. For the landscape visual impact, the overlap analysis determines the overlap between wind farms and the ternary scenicness classification, excluding "scenic", and "highly scenic" areas ("other" is never excluded) if less than 15% of existing wind farms overlap with the respective category. Further details on this analysis and the resulting exclusions are shown in Supplementary Note 5 and Table S5.5. The buffer distance from coastlines is set to 12 nautical miles (22,224m), in line with the medium restrictions from Ref.[25].

For the low level of tolerance to social impacts, regulatory setback distances to settlements are set to 3,000m, corresponding to the highest value reported in the literature[77]. Shadow flicker impacts are eliminated completely (or at least limited to a maximum of 1 minute of exposure per year) and set to 2,500m, while the setback distance due to noise is set to 2,000m, corresponding to the maximum value of Ref.[78]. All scenic and highly scenic areas are excluded across Europe, and the distance to the coastline is set to 12 nautical miles (22,224m).

*Technical restrictions*

In addition to the social and environmental impact tolerance scenarios, we exclude land area from wind energy deployment (both onshore and offshore) based on more technical factors. For example, installing wind turbines at or close to existing infrastructure such as airports, roads and military areas is not feasible. Such exclusions are considered technical and are applied uniformly across all scenarios. The categories considered and eventual buffer distances are shown in Supplementary Table S3.2 and include most of the categories identified by Ref.[77] that are considered relevant for the geographical potential of onshore wind energy. The technical exclusions amount to 44.2% of the total EU25+3 area and a geospatial overview of the remaining land-area is shown in Supplementary Figure S3.1.

**Soft-coupling of JRC-EU-TIMES and highRES-Europe**

Two energy system models are used to generate our results, the whole-energy system model JRC-EU-TIMES and the electricity system model highRES-Europe. JRC-EU-TIMES provides insights on the transition of the full European energy system to net-zero (see Methods - JRC-EU-TIMES modelling setup for further details), whereas highRES-Europe models the electricity system in high spatial and temporal detail (see Methods - highRES-Europe modelling setup for further details). The models are soft-coupled and Supplementary Figure S4.1 outlines the linkage between the two models and their interactions.

JRC-EU-TIMES optimises the transition to a future net-zero European energy system, complete with detailed policy representation, and provides 2050 boundary conditions for highRES-Europe. Key aspects included are country-level electricity demands and $CO_2$ emission targets for the electricity system that are aligned with a net-zero whole energy system. Supplementary Table S4.2 shows the annual electricity demand and $CO_2$ emission budget for the EU25+3 countries modelled in 2050.

All permutations of the social and environmental tolerance scenarios combined with the technical restrictions are then used to exclude land area from wind energy deployment. The remaining land area is converted into capacity potentials per NUTS2 using 2.4 MW/km$^2$ for onshore (see Supplementary Note 4 and Supplementary Table S4.6) and 5 MW/km$^2$ for offshore. These potentials, in conjunction with the whole-energy system boundary conditions, feed highRES-Europe, which optimises the installed capacity and spatial deployment of electricity, storage, and transmission infrastructure as well as hourly operations for the European electricity system in 2050 across all nine scenarios.

**JRC-EU-TIMES modelling setup**

JRC-EU-TIMES[81] is a linear optimisation, bottom-up, partial equilibrium energy system model developed within the IEA-ETSAP TIMES framework[82]. It is open source and underpins several EU-wide energy system models within the ETSAP community. The model is designed to analyse long-term transitions of the European energy system under alternative technology, resource and policy assumptions. It minimises the total discounted energy system cost while meeting exogenously specified energy service demands and policy constraints, including greenhouse gas (GHG) emission limits. Total system costs include investments, fixed and variable operation and maintenance costs, fuel expenditures and decommissioning costs. The model represents the entire energy system from primary resource supply and imports through transformation, transmission and distribution to final energy demand. End-use sectors include seven industrial sectors, six services subsectors, detailed residential building categories and 18 transport modes.

In this application, JRC-EU-TIMES covers the EU27 plus the United Kingdom, Norway and Switzerland (EU27+3) at country-level resolution. The time horizon extends to 2050 in five-year steps. Each model year is divided into 12 intra-annual time slices (day, night and peak across four seasons). In the power sector, time slices distinguish conditions with and without surplus variable renewable energy (VRE), enabling endogenous representation of curtailment, storage and sector coupling within an aggregated temporal structure. Wind technologies are disaggregated into 14 onshore and offshore categories differentiated by capacity factor ranges to better capture resource heterogeneity.

To ensure net-zero GHG emissions and compliance with EU climate neutrality by 2050, this application of JRC-EU-TIMES comprehensively represents relevant European legislation and policy instruments through GHG caps, renewable and efficiency targets as well as sector-specific constraints. Full details of the relevant policy instruments and how they are operationalised are available in Supplementary Note 4 and Supplementary Table S4.1.

**highRES-Europe modelling setup**

The high spatial and temporal resolution electricity system model for Europe (highRES-Europe) is a linear optimisation electricity model framework, specifically designed to represent electricity systems with a high share of variable renewable electricity generation[83]. The model optimises (minimises) the total system cost (annualised investment and operational costs) while ensuring that supply and demand is balanced at an hourly level and that technical and operational constraints are met. The total system costs are minimised both by spatially detailed investment decisions and the operation (dispatch) of the system for a snapshot year (here 2050).

While the model framework is flexible, this application includes a spatial coverage of EU25 + 3 (Norway, Switzerland and the United Kingdom). Demand and transmission are balanced at the country level, but capacity deployment for wind- and solar power is made at a finer resolution (NUTS2 level). The available generation and storage technologies in our Base scenario, i.e. the case which underpins all the of the results reported in the main manuscript, are shown in Supplementary Table S4.3. As in Ref.[84], linearised unit commitment constraints are used to represent the operational details of fossil fuel, biomass, hydrogen and hydropower generation. These simplifications are made to maintain computational tractability.

*Hourly VRE time series and bias correction*

To generate the hourly capacity factor time series at a NUTS2 resolution, which are used by highRES, we use atlite[85] to convert ERA5 reanalysis[86] weather data (e.g. wind speeds and solar irradiance) into energy systems data. The spatial and temporal resolution is set by the input data, which for ERA5 is hourly data on a 0.25ºx0.25º grid. While ERA5 is the standard dataset for weather variables, it is well known to hold some bias, particularly in regions with complicated topography, such as the Alps and the Scandinavian Mountains[87]. To mitigate this, we bias-correct the 100m wind speed from ERA5 using correction ratios[88] derived by spatially averaging microscale values from the Global Wind Atlas[89] at 0.025x0.025º and dividing them by the corresponding ERA5 values. See Supplementary Note 4 for further details and Supplementary Figure S4.3 for the effect of the bias-correction. An overview of the resulting average annual capacity factors for the high-high scenario for each NUTS2 region is provided in Supplementary Figure S4.4 to give an idea of what the (relatively) unconstrained model is given.

*Biomass potentials and BECCS*

The biomass potentials in highRES-Europe are the medium scenario taken from JRC ENSPRESO[90]. We exclude: i) energy crops to minimise competition with food production, ii) imports to ensure the highest standards for feedstock sustainability and iii) primary forest residues to mitigate concerns around the sustainability of that supply chain. This results in a total biomass potential for the 28 countries modelled of 1,000 TWh/yr.

The boundary conditions from our whole energy system modelling include the requirement for certain countries to achieve net negative emissions from the electricity sector by 2050. To achieve this, we model bioenergy with carbon capture and storage (BECCS) which, following Ref.[91], we assume can only provide a limited amount of compensation for concurrent fossil fuel emissions. This is taken here to be a maximum sequestration of 5 $MtCO_2$/yr beyond what is needed to achieve the emissions target coming from the whole energy system model for the combined electricity system of all 28 countries (~ -81 $MtCO_2$/yr). This choice is made to prioritise emissions reductions rather than offsetting via BECCS and also to reflect the substantial array of non-modelled factors surrounding speculative technology options like BECCS, such as risks and uncertainties around scale up and wider social, environmental and economic impacts[92,93]. We relax this assumption in one of the sensitivities we describe below.

*Technology data and restrictions*

highRES-Europe includes technical constraints as well as assumptions on economic input parameters, such as technology cost parameters, which are shown in Supplementary Table S4.4. The topology of cross-border transmission follows the reference grid of ENTSO-E Ten-Year-Network-Development-Plan (TYNDP)[94] with investments in capacity reinforcement by 2050 limited to the potential network candidates listed in that report. Intra-country transmission is modelled as a copperplate.

Whereas the capacity potential for onshore wind is based on the social and environmental tolerance scenarios, the available land area for solar PV is restricted based on some common exclusion criteria[95], including areas with a steep slope (> 6.3°)[96], areas with medium and high agricultural intensity crops and grasslands[97], CDDA categories Ia-IV[71], Natura 2000 areas[72] as well as CORINE land cover types[73] (urban areas, forests, semi-natural areas, wetlands and water bodies). Rooftop areas for PV are taken from Ref.[98] with the area for countries not in that dataset based on extrapolation of per capita roof area for similar countries, i.e. Sweden for Norway, Slovenia for Croatia and Austria for Switzerland. The total rooftop area per country is downscaled to the NUTS2 level by proportionally sharing it out based on the total footprint area of CORINE classes continuous urban fabric, discontinuous urban fabric and industrial and commercial (codes 1, 2 and 3) in each NUTS2 to the total area of these classes in each country. This NUTS2 level rooftop area is then added to the land area available for PV.

In addition to land availability restrictions, solar PV deployment is also restricted by techno-economic factors. Build-out rates may be limited due to, for example, material scarcity, grid integration or financial profitability[99]. To account for this, and constrain technology deployment in 2050, we limit upper capacity deployment for each of the 28 countries in the model. The limit is based on a linear extrapolation from 2024 to the target year (2050), with a growth rate equal to 1.5 times the historically highest annual capacity deployment for each country and technology individually. Details on the process and results are shown in Supplementary Note 4 and Supplementary Table S4.5.

*Existing capacities*

We model existing wind, solar, and nuclear capacities that are still online in 2050 in highRES-Europe using Global Energy Monitor (GEM) data[100] for the 28 European countries included in this study. We estimate each country's and NUTS2 region's 2050 capacity by including all plants projected to still be in operation in 2050 based on their expected decommissioning dates. We estimate this using the commissioning dates provided by GEM and lifetimes for each technology: 30 years for wind and solar, and 40 years for nuclear, inferred from GEM data. Additional details about the capacity calculation are available in the Supplementary Note 4 and Supplementary Figure S4.5.

We also assume all of today's run-of-river and reservoir hydropower capacity together with pumped hydro found in the EU25+3 countries we model is still online in 2050[101].

*Sensitivity runs*

To explore the impact of a number of critical assumptions on our results, we have modelled five sensitivities (*BECCS+, $H_2U$, Trans+, PV+* and *Floating*) for all nine tolerance scenarios. The details of how the sensitivity scenarios are implemented, and their rationale is included in Supplementary Note 2 and Supplementary Table S2.1. In short, *BECCS+* allows for additional $CO_2$ sequestration from BECCS and that negative emissions can used to offset contemporaneous fossil fuel combustion. *$H_2U$* includes long duration hydrogen storage in salt caverns. *Trans+* allows further expansion of interconnectors in the TYNDP to three times the capacity limit used in our Base scenario. *PV+* removes the growth rate constraint on solar PV deployment, meaning that it is only constrained by land availability. *Floating* introduces floating offshore wind as a new technology, opening up marine areas that were previously unavailable due to depths over 65 metres.

**Data and code availability**

The version of the highRES-Europe model workflow used in this study is openly available on GitHub (https://github.com/highRES-model/highRES-Europe-WF/tree/WIMBY), as is the GAMS code (https://github.com/highRES-model/highRES-Europe-GAMS/tree/wimby_gams). Input data required for running highRES-Europe and replicating the results is available on Zenodo (https://doi.org/10.5281/zenodo.19087632). This includes both default data used in highRES-Europe and adaptations for this study. Additionally, the Python-based scripts used to generate the land availability exclusions, demand data and existing capacities are available in a separate preprocessing GitHub repository (https://github.com/highRES-model/highRES-Europe-PreProc-WF/).

**Acknowledgements**

This research has received funding from the European Union's Horizon Europe research and innovation programme under grant agreement no. 101083460 (WIMBY). We also thank Steve Pye for his feedback on the manuscript.

**Author contributions**

**James Price**: Conceptualisation, Methodology, Software, Formal analysis, Investigation, Data curation, Writing – Original draft, Visualisation, Supervision, Funding acquisition.
**Guillermo Valenzuela-Venegas**: Conceptualisation, Methodology, Software, Formal analysis, Data curation, Writing – Original draft, Visualisation.
**Oskar Vågerö**: Conceptualisation, Methodology, Software, Formal analysis, Data curation, Writing – Original draft, Visualisation.
**Marianne Zeyringer**: Conceptualisation, Methodology, Formal analysis, Data curation, Writing – Original draft, Supervision, Funding acquisition.

**Meixi Zhang**: Conceptualisation, Methodology, Investigation, Writing – Review & editing.
**Evangelos Panos**: Conceptualisation, Methodology, Investigation, Writing – Review & editing.
**Ruihong Chen**: Conceptualisation, Methodology, Formal analysis, Writing – Review & editing.
**Adrienne Etard**: Conceptualisation, Methodology, Formal analysis, Writing – Review & editing.
**Andrea N. Hahmann**: Conceptualisation, Methodology, Writing – Review & editing, Funding acquisition.
**Luis Ramirez Camargo**: Conceptualisation, Methodology, Writing – Review & editing, Project administration, Funding acquisition.
**Alena Lohrmann**: Conceptualisation, Methodology, Writing – Review & editing.
**Russell McKenna**: Conceptualisation, Writing – Review & editing, Funding acquisition.
**Christian Mikovits**: Conceptualisation, Methodology, Formal analysis, Writing – Review & editing, Funding acquisition.
**Monika Bucha**: Conceptualisation, Methodology, Formal analysis, Investigation, Writing – Review & editing.

**Competing interests**
The authors declare no competing interests

**ORCIDs**
Oskar Vågerö: https://orcid.org/0000-0002-0806-0329
Adrienne Etard: https://orcid.org/0000-0002-1700-2972
Andrea Hahmann: https://orcid.org/0000-0001-8785-3492
Luis Ramirez Camargo: https://orcid.org/0000-0002-1554-206X
Alena Lohrmann: https://orcid.org/0000-0003-3763-1415
Christian Mikovits: https://orcid.org/0000-0001-6048-1916
Guillermo Valenzuela-Venegas: https://orcid.org/0000-0003-2067-7257

# Supplementary information for “The electricity system value of the local acceptance of onshore wind in Europe”

# Supplementary information

## Supplementary note 1 – Detailed onshore wind energy potentials and results

In this section we provide additional spatial details to the onshore wind energy potentials in Figure 2. While the results are the same, the following figures show particular regions and simplifies visual inspection of the results. Data on exact onshore wind energy potentials for each modelled NUTS2 region is available in Supplementary Data File 1.

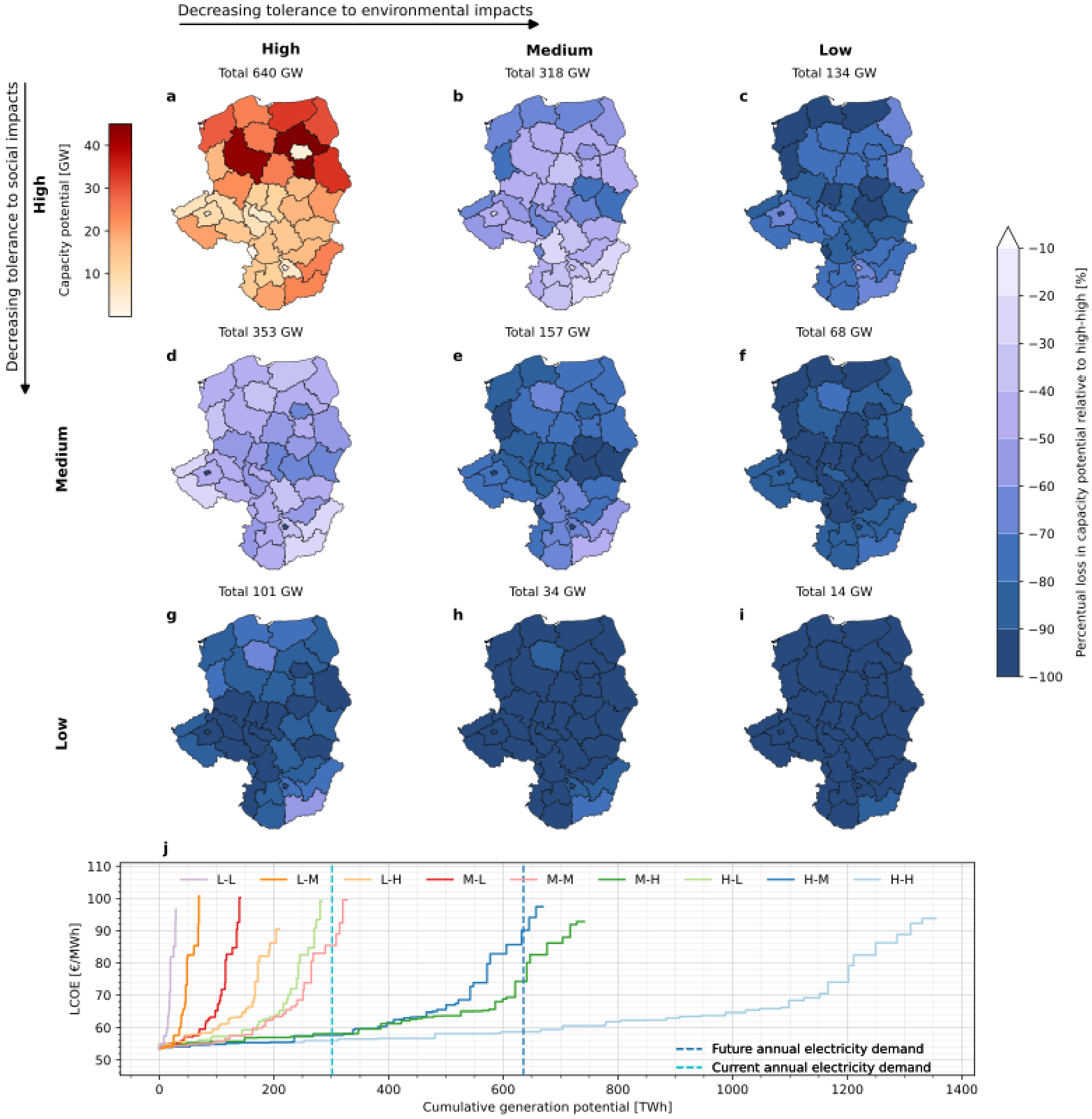

**Figure S1.1 Eligible capacity potential for wind energy deployment across nine scenarios a-i) of tolerance to social and environmental impacts for Eastern Europe (PL, CZ, SK and HU).**

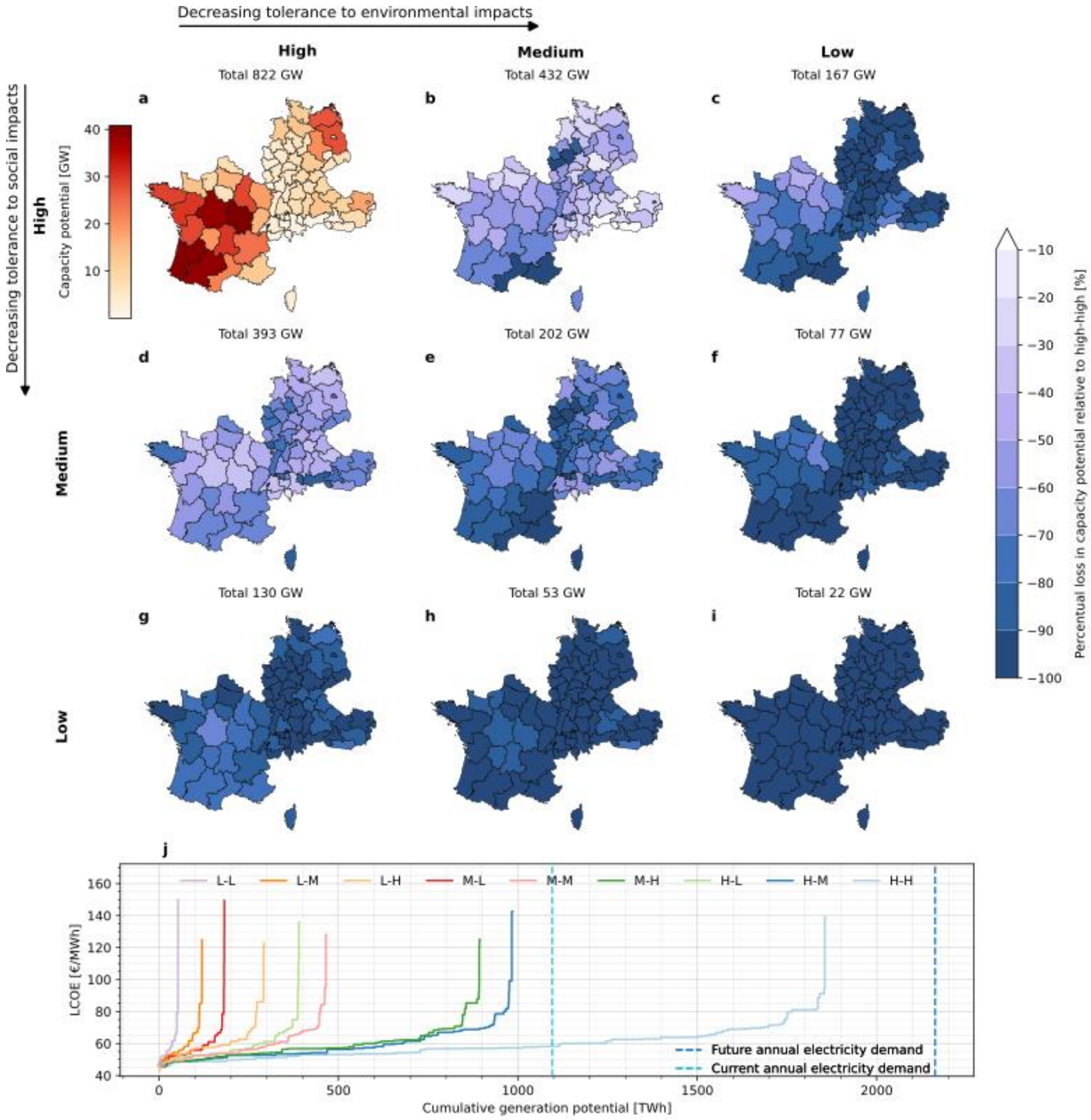

**Figure S1.2 Eligible capacity potential for wind energy deployment across nine scenarios a-i) of tolerance to social and environmental impacts for Western Europe (DE, FR, AT and CH).**

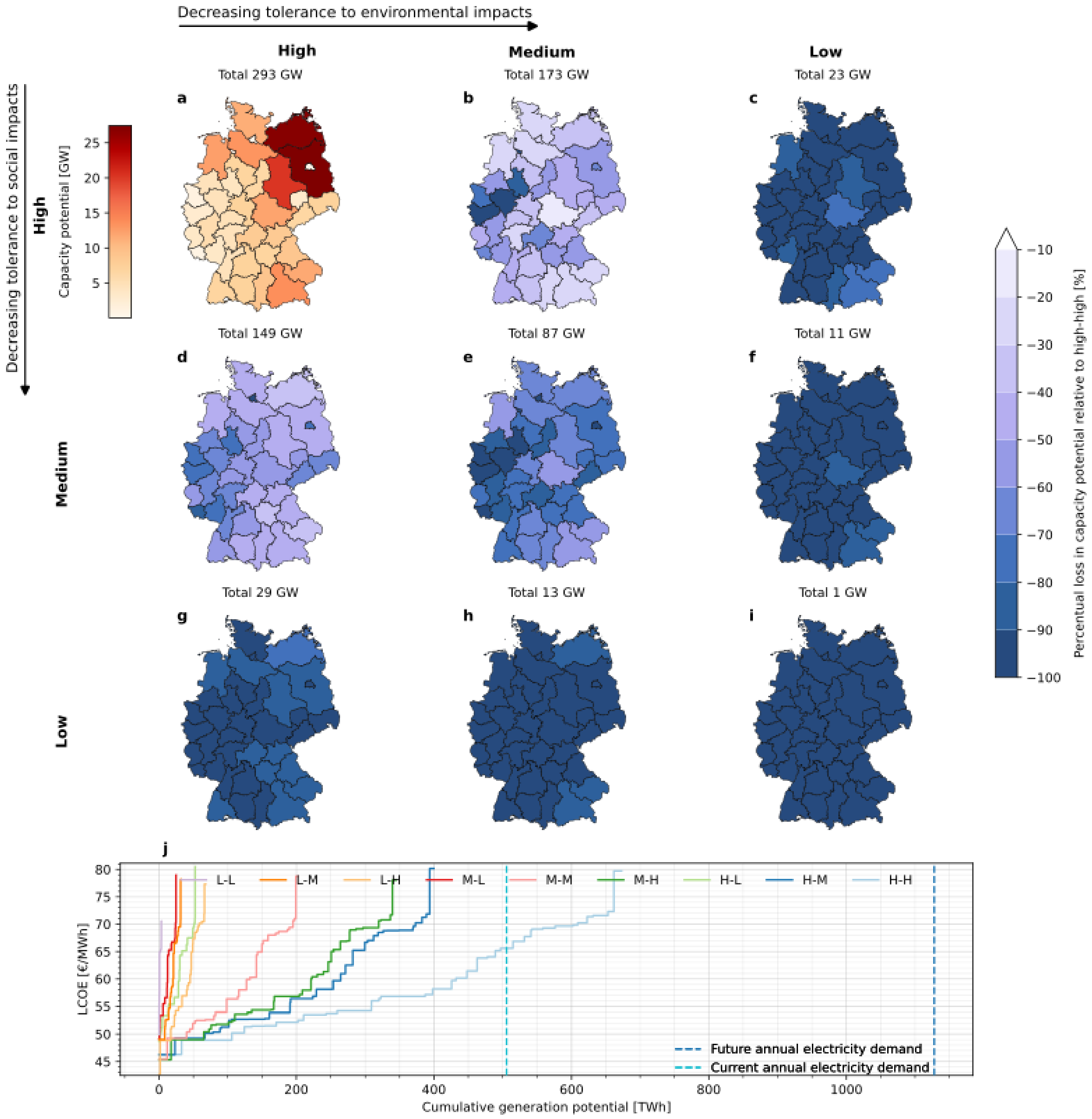

**Figure S1.3 Eligible capacity potential for wind energy deployment across nine scenarios a-i) of tolerance to social and environmental impacts for Germany (DE).**

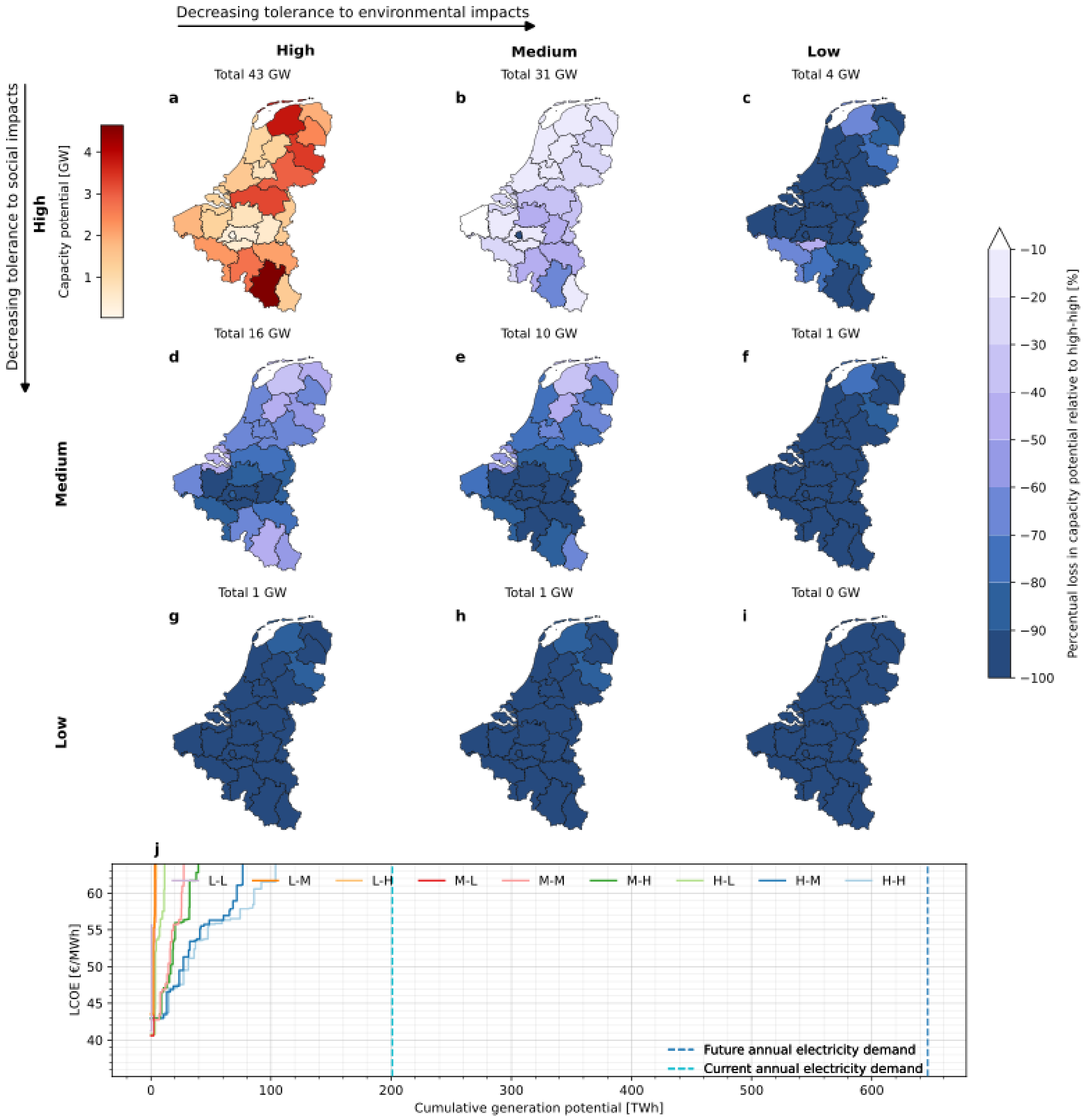

**Figure S1.4 Eligible capacity potential for wind energy deployment across nine scenarios a-i) of tolerance to social and environmental impacts for Benelux (BE, NL, LU).**

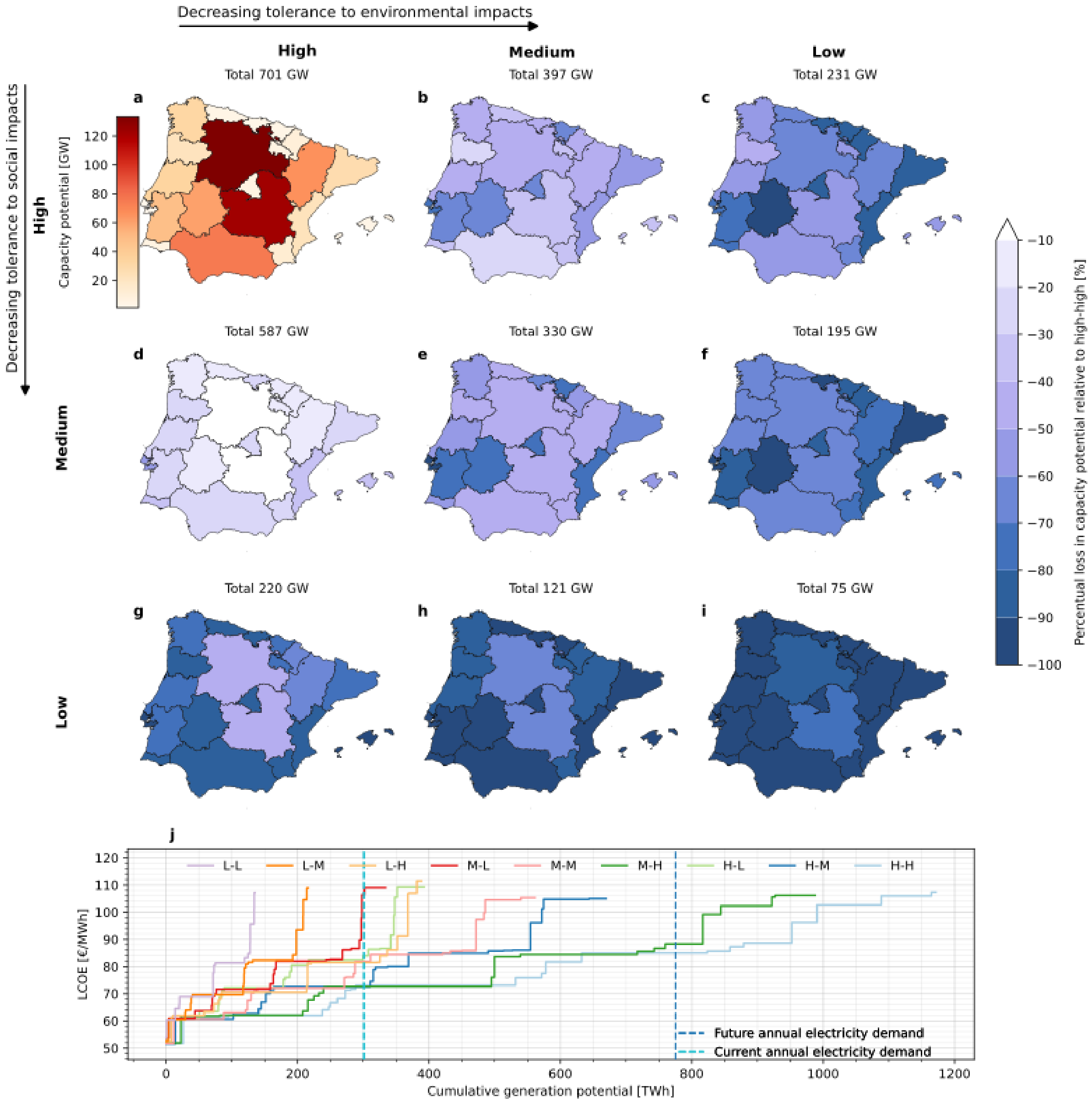

**Figure S1.5 Eligible capacity potential for wind energy deployment across nine scenarios a-i) of tolerance to social and environmental impacts for the Iberian Peninsula (ES, PT).**

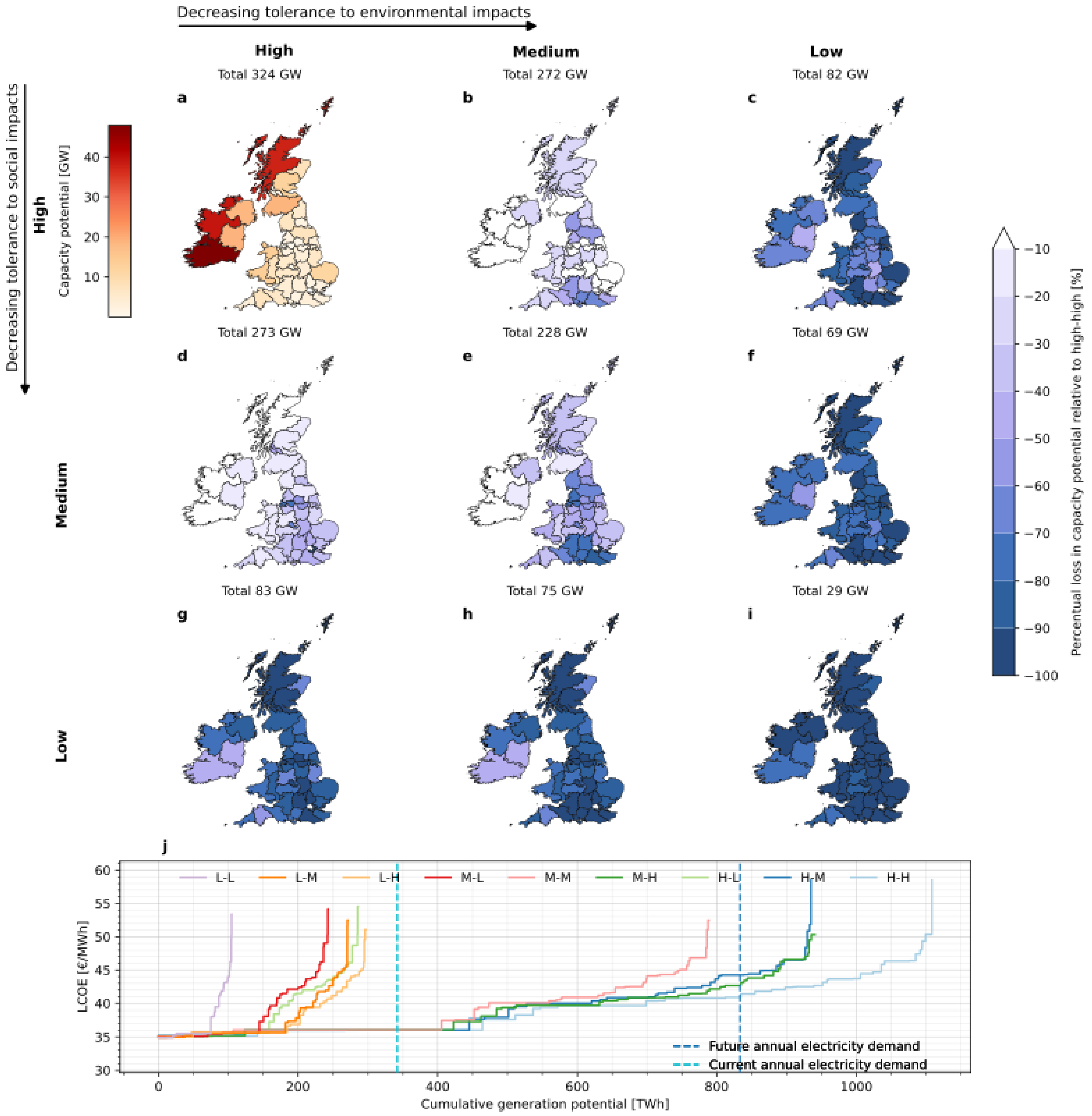

**Figure S1.6 Eligible capacity potential for wind energy deployment across nine scenarios a-i) of tolerance to social and environmental impacts for Ireland and the United Kingdom (IE, UK).**

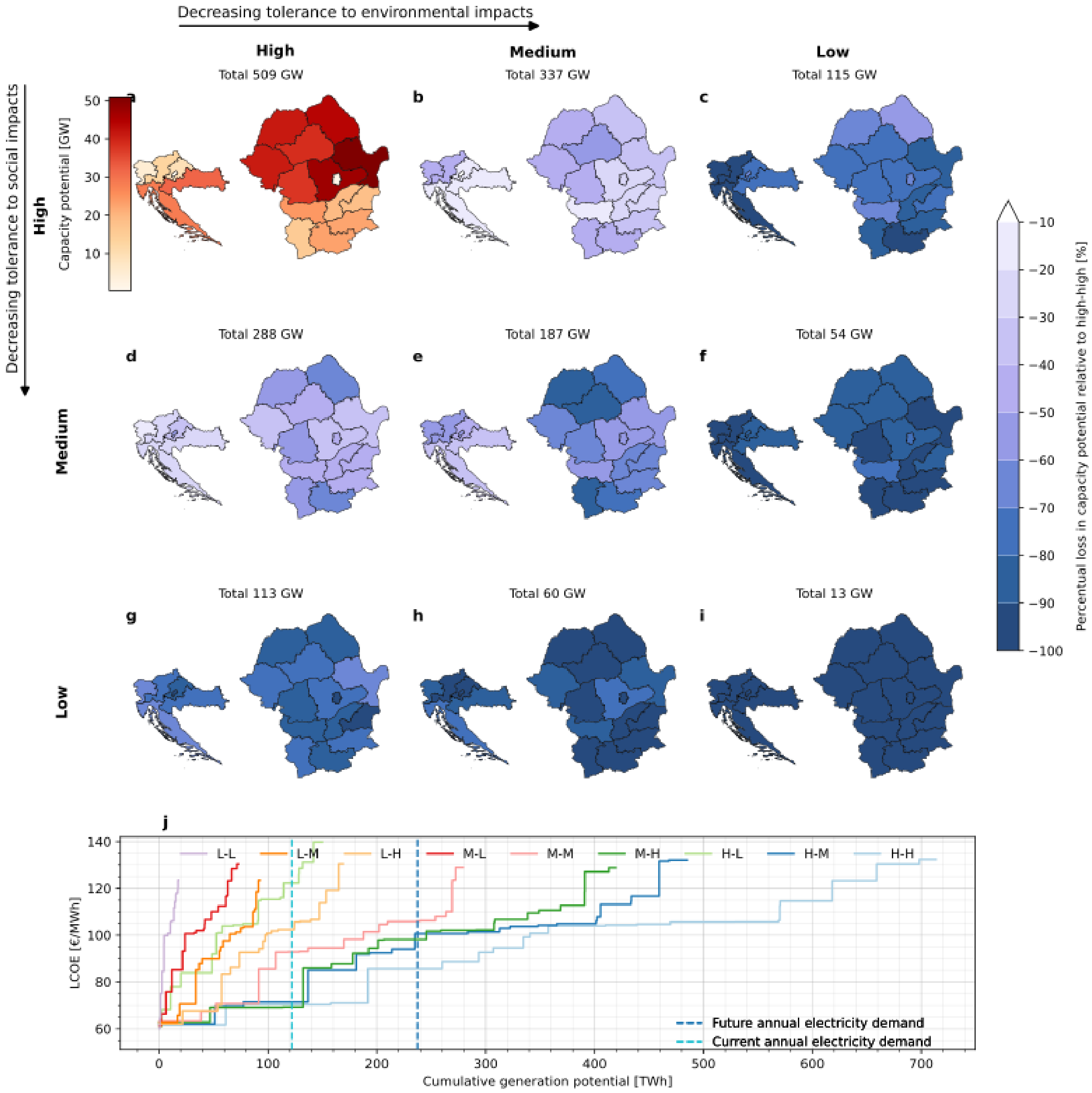

**Figure S1.7 Eligible capacity potential for wind energy deployment across nine scenarios a-i) of tolerance to social and environmental impacts for some of the countries on the Balkan Peninsula (BG, HR, RO, SI).**

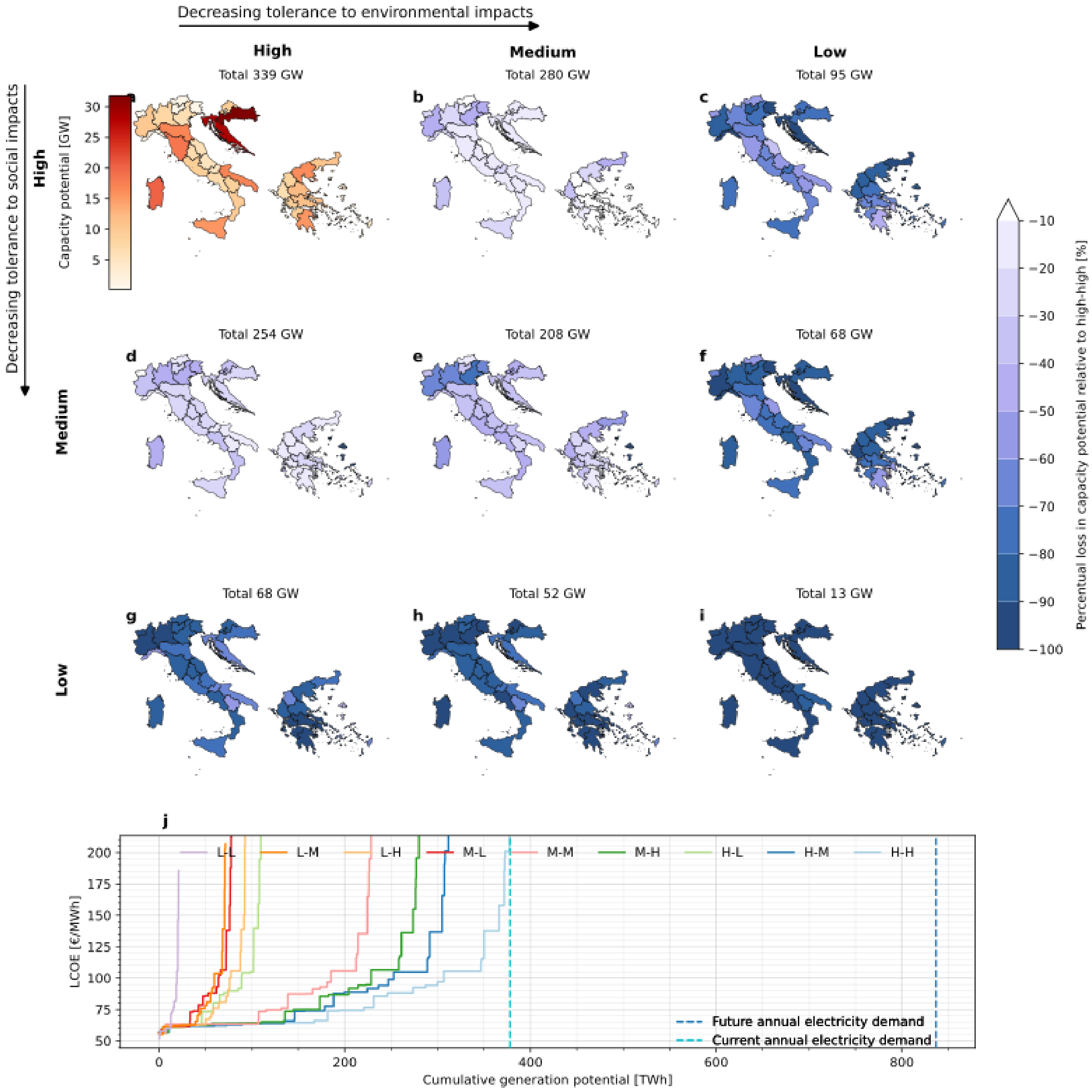

**Figure S1.8 Eligible capacity potential for wind energy deployment across nine scenarios a-i) of tolerance to social and environmental impacts for countries at the Adriatic Sea (HR, GR, IT).**

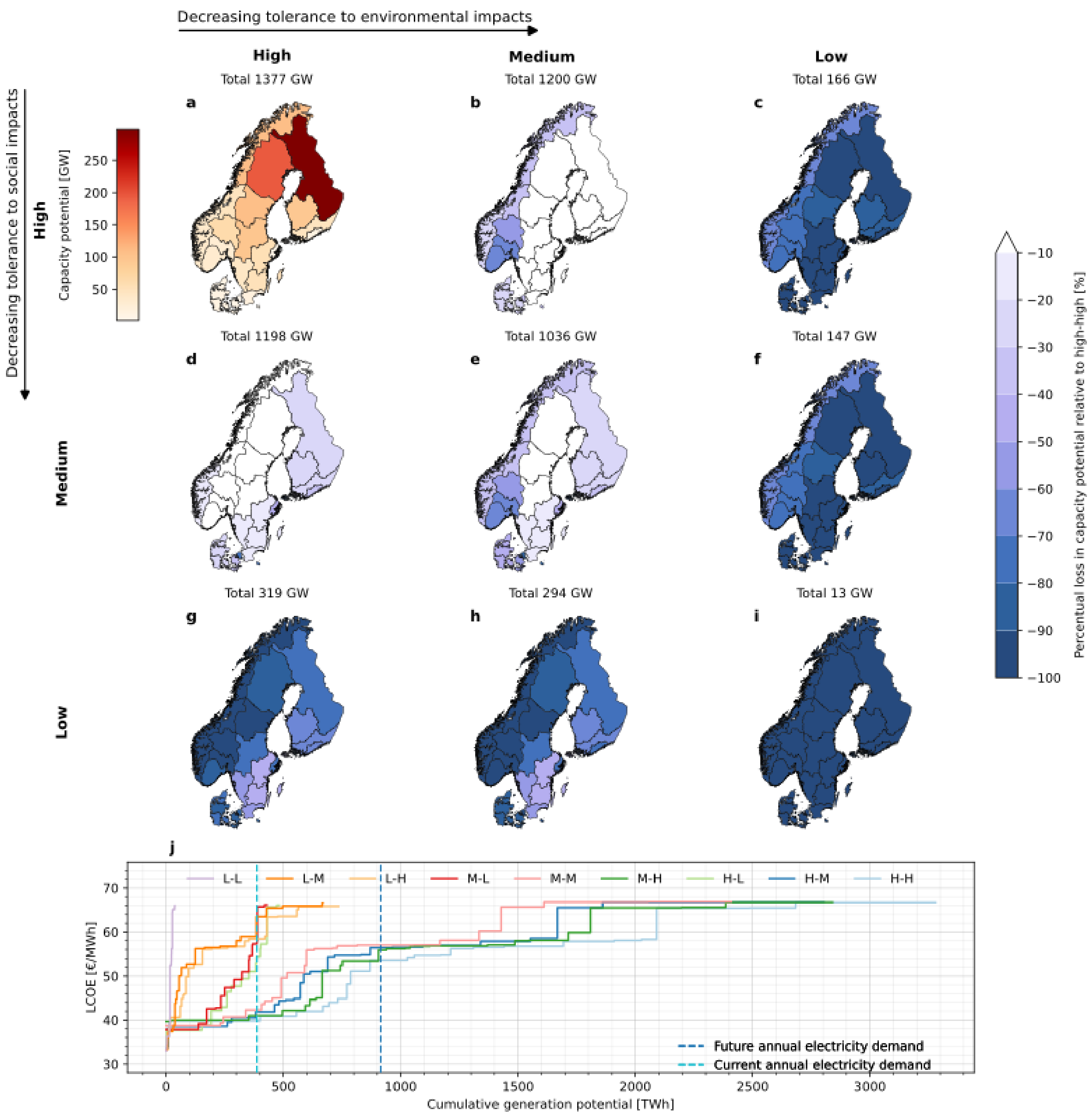

**Figure S1.9 Eligible capacity potential for wind energy deployment across nine scenarios a-i) of tolerance to social and environmental impacts for the modelled Nordic countries (DK, FI, NO, SE).**

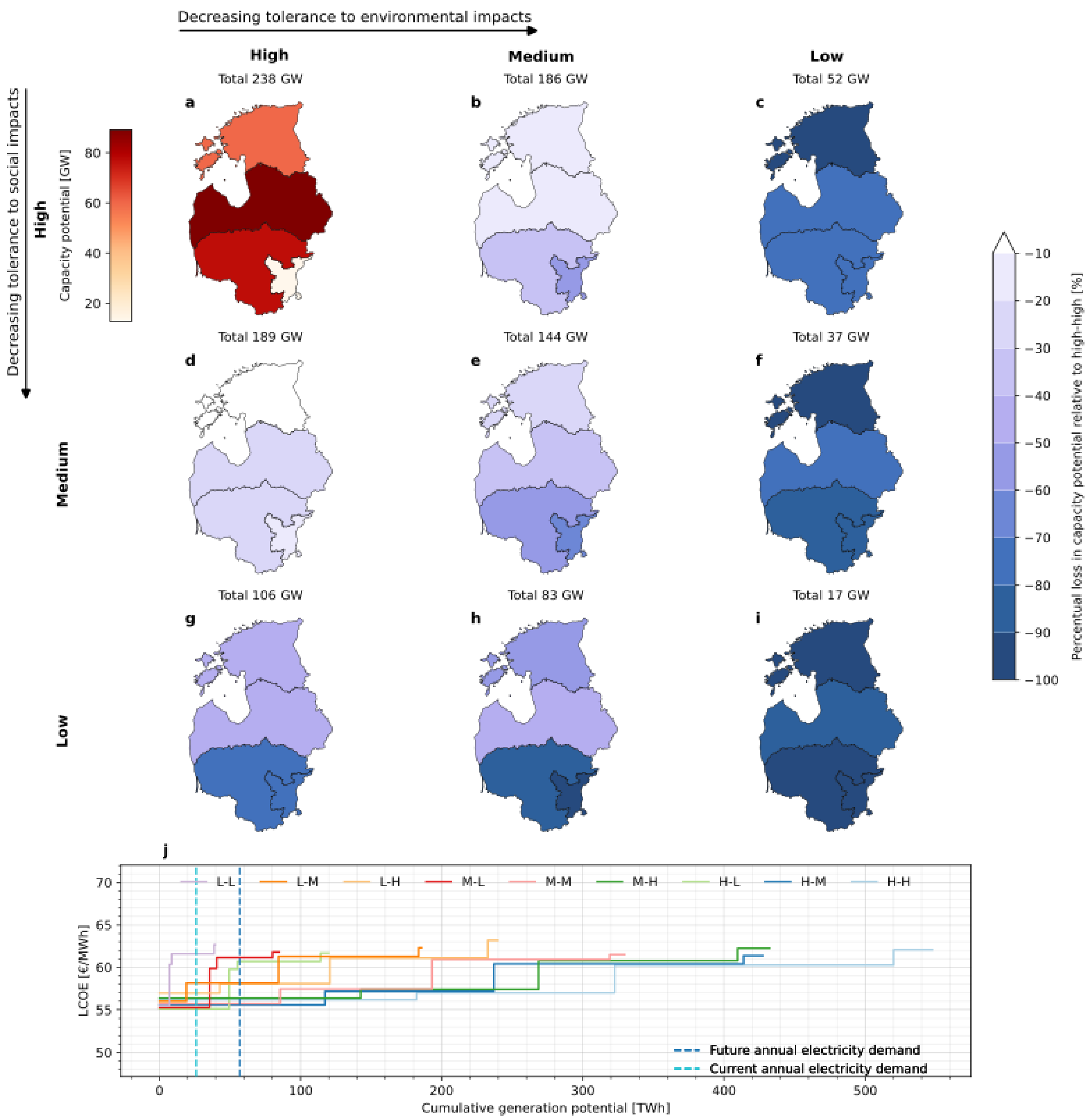

**Figure S1.10 Eligible capacity potential for wind energy deployment across nine scenarios a-i) of tolerance to social and environmental impacts for the Baltic countries (EE, LV, LT).**

**Table S1.1 Onshore wind energy potential (GW) on the country-level across the nine scenarios.**

| Country | H-H | H-M | H-L | M-H | M-M | M-L | L-H | L-M | L-L |
|---|---|---|---|---|---|---|---|---|---|
| Austria | 46.3 | 36.1 | 4.3 | 16.5 | 13.0 | 1.2 | 4.1 | 2.8 | 0.3 |
| Belgium | 16.8 | 10.2 | 2.2 | 5.1 | 1.8 | 0.2 | 0.4 | 0.0 | 0.0 |
| Bulgaria | 122.0 | 85.3 | 22.7 | 59.8 | 40.5 | 12.1 | 22.6 | 11.3 | 3.4 |
| Czechia | 71.9 | 33.2 | 18.2 | 43.0 | 16.7 | 9.4 | 7.0 | 2.0 | 1.2 |
| Croatia | 69.6 | 60.9 | 11.0 | 52.5 | 45.0 | 7.3 | 19.0 | 15.1 | 1.2 |
| Denmark | 42.3 | 31.2 | 0.0 | 29.5 | 21.5 | 0.0 | 8.5 | 5.8 | 0.0 |
| Estonia | 59.9 | 51.3 | 2.7 | 54.2 | 46.3 | 2.3 | 33.8 | 28.4 | 0.7 |
| Finland | 452.7 | 443.1 | 38.9 | 343.5 | 336.2 | 28.9 | 135.0 | 131.9 | 8.1 |
| France | 474.2 | 217.5 | 138.6 | 222.8 | 99.0 | 64.4 | 96.6 | 36.0 | 21.0 |
| Germany | 293.7 | 173.6 | 23.6 | 149.2 | 87.2 | 11.2 | 29.5 | 14.0 | 1.6 |
| Greece | 98.7 | 84.2 | 27.6 | 78.1 | 66.3 | 21.3 | 15.5 | 12.2 | 1.8 |
| Hungary | 115.2 | 75.4 | 34.3 | 71.6 | 45.9 | 21.5 | 26.2 | 15.4 | 7.1 |
| Ireland | 105.9 | 105.9 | 35.3 | 97.6 | 97.6 | 32.8 | 45.9 | 45.9 | 18.6 |
| Italy | 171.6 | 135.4 | 57.3 | 123.6 | 96.9 | 40.3 | 34.4 | 25.3 | 10.1 |
| Latvia | 89.2 | 80.1 | 26.7 | 64.2 | 57.5 | 18.2 | 51.2 | 45.7 | 13.9 |
| Lithuania | 89.5 | 55.2 | 23.1 | 70.8 | 40.6 | 16.8 | 21.6 | 9.0 | 3.3 |
| Luxembourg | 1.4 | 1.1 | 0.1 | 0.6 | 0.5 | 0.0 | 0.0 | 0.0 | 0.0 |
| Norway | 314.0 | 178.8 | 96.5 | 291.6 | 165.2 | 88.7 | 23.9 | 10.2 | 5.0 |
| Netherlands | 24.8 | 19.7 | 2.2 | 10.4 | 8.4 | 1.2 | 1.5 | 1.2 | 0.2 |
| Poland | 408.3 | 181.5 | 75.3 | 212.9 | 80.6 | 34.0 | 63.6 | 15.7 | 6.0 |
| Portugal | 111.1 | 55.2 | 39.3 | 84.0 | 42.0 | 29.4 | 20.6 | 10.5 | 6.6 |
| Romania | 300.3 | 181.6 | 81.3 | 161.7 | 93.9 | 34.3 | 67.5 | 31.9 | 8.4 |
| Spain | 590.9 | 342.3 | 192.2 | 503.6 | 288.9 | 166.3 | 199.5 | 111.4 | 69.2 |
| Slovenia | 17.4 | 10.1 | 1.0 | 14.0 | 7.9 | 0.8 | 4.5 | 1.9 | 0.1 |
| Slovakia | 44.7 | 28.3 | 6.8 | 25.8 | 14.6 | 3.2 | 4.2 | 1.6 | 0.2 |
| Sweden | 569.0 | 547.6 | 31.4 | 533.7 | 513.9 | 29.9 | 152.2 | 146.1 | 0.7 |
| Switzerland | 8.7 | 5.4 | 1.1 | 5.3 | 3.4 | 0.8 | 0.7 | 0.4 | 0.1 |
| United Kingdom | 218.8 | 166.6 | 47.2 | 175.6 | 130.4 | 36.7 | 37.4 | 29.8 | 10.4 |
| EU25+3 | 4,928.8 | 3,396.9 | 1,040.6 | 3,501.2 | 2,461.8 | 713.2 | 1,126.9 | 761.5 | 199.4 |

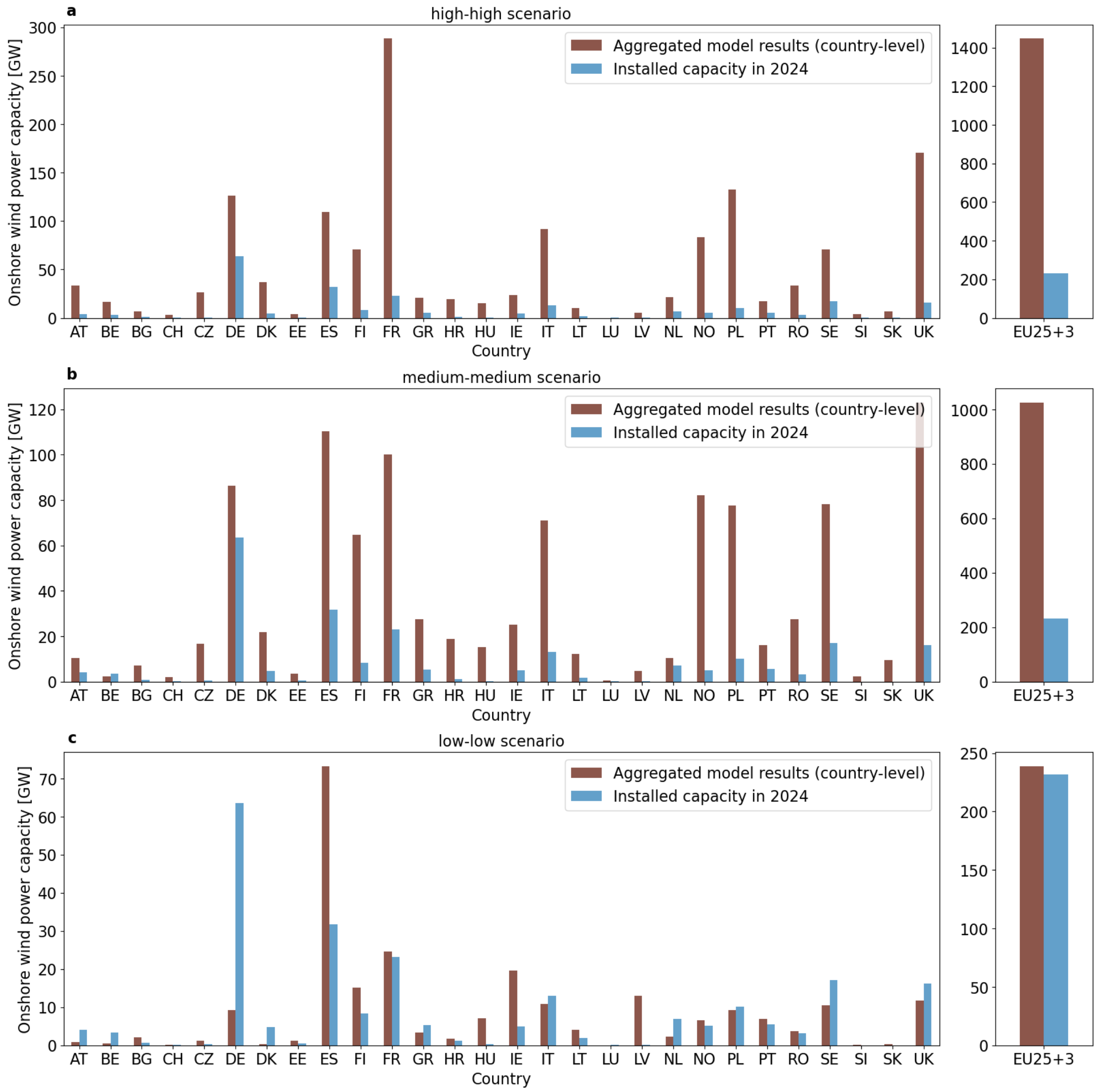

**Figure S1.11 Comparison between the diagonal of the modelling results to current (2024) onshore wind energy deployment[1] on the country level and for EU25+3**. Model results show that Europe moves from a considerable deployment, exceeding the existing onshore wind capacity seven times over (high-high scenario, panel a), to a more moderate deployment (medium-medium scenario, panel b) and ultimately only slightly exceed the existing onshore wind capacity (low-low scenario, panel c).  Note that the y-axis scale changes between panel a-c.

# Supplementary note 2 – Sensitivity analysis

In this section we explore the implications for our results of varying various aspects of our electricity system modelling set-up away from their assumptions in our core Base scenario in highRES-Europe. Specifically, we look at a set of additional sensitivities as detailed in Table S2.1.

**Table S2.1 highRES-Europe sensitivity scenarios.**

| Sensitivity | Main adjustment |
|---|---|
| BECCS+ | In this case we remove the 5 $MtCO_2$ limit on sequestration from BECCS beyond that required to achieve the overall emissions target specified for the electricity system by JRC-EU-TIMES. This means that negative emissions from BECCS can be used to offset contemporaneous fossil fuel combustion. |
| $H_2U$ | In addition to all the technologies available in Base we also make hydrogen long duration storage in salt caverns available to the electricity system optimisation. Underground long duration storage is an emerging technology option with large potential but limited deployment as of yet, hence it was not included in our Base scenario configuration. |
| Trans+ | We permit interconnection between countries to be expanded up to three times more than the limits imposed in Base which are taken from ENTSO-E's Ten-Year Network Development plan from 2024[2]. This is done to explore how much larger amounts of interconnection expansion alters our findings. |
| PV+ | Solar PV deployment in each country is no longer limited based on 1.5 times the historical maximum growth rate from IRENA[1]. This means that solar PV deployment is only restricted by land availability, allowing for much higher than historic deployment rates and a greater role for solar PV. |
| Floating | In addition to all technologies available in Base we also make floating offshore wind an option in the highRES-Europe optimisation. This is an emerging technology with limited deployment to date and so was not included in our Base scenario. |

In Fig S2.1 we plot the system levelised cost of electricity (LCOE) for the whole of Europe modelled here in our sensitivity cases and our Base scenario configuration (left panel) and how it changes across the land availability scenarios (right panel). A general pattern of increasing system costs as the role of onshore wind is constrained is found for all cases, with medium-medium seeing a 3-4% increase and low-low a 10-15% uptick. The BECCS+, PV+ and Floating sensitivities show the smallest growth in costs, with a maximum of ~10-11%, implying they are least impacted by a heavily limited role for onshore wind.

In BECCS+, the European electricity system pivots to a greater reliance on natural gas generation to more cost-effectively compensate for the loss of onshore wind with the associated extra emissions offset by greater carbon dioxide removal. Carbon dioxide sequestration levels are ~400 $MtCO_2$/yr (BECCS and fossil CCS combined), over three times that of the Base scenario (~130 $MtCO_2$/yr). These high levels of sequestration may not be feasible[3], particularly given that other sectors within the whole energy system will also utilise sequestration capacity. Greater fossil fuel consumption would also mean a greater exposure for Europe to volatile international fossil fuel markets and the associated energy security challenges that brings.

PV+ would see Europe's electricity system deploying more solar PV capacity in 2050 than permitted in Base to compensate for onshore wind being constrained. In the most pessimistic case (low-low) for wind an additional nearly 750 GW of solar PV is installed compared to Base which, if all ground mounted, would use an extra ~18,750 km2 of land. This sensitivity would also require a number of countries to vastly exceed their historic peak installation rates for solar PV (based on data from IRENA[1]). For example, in Italy and Spain the average annual build-out rate would need to increase by around 300% and France would require an even faster increase in deployment by over 500%.

The Floating sensitivity has floating offshore wind deployed in progressively larger volumes to compensate for onshore wind becoming more and more limited. The technology is particularly important for countries like Italy and Spain which have limited bottom mounted offshore wind potentials due to the interactions between water depth and spatial buffers that limit deployment too close to coastlines in our scenarios. In low-low, Europe deploys ~265 GW of floating wind, nearly the EU's policy target for the entire offshore wind sector by 2050[4].

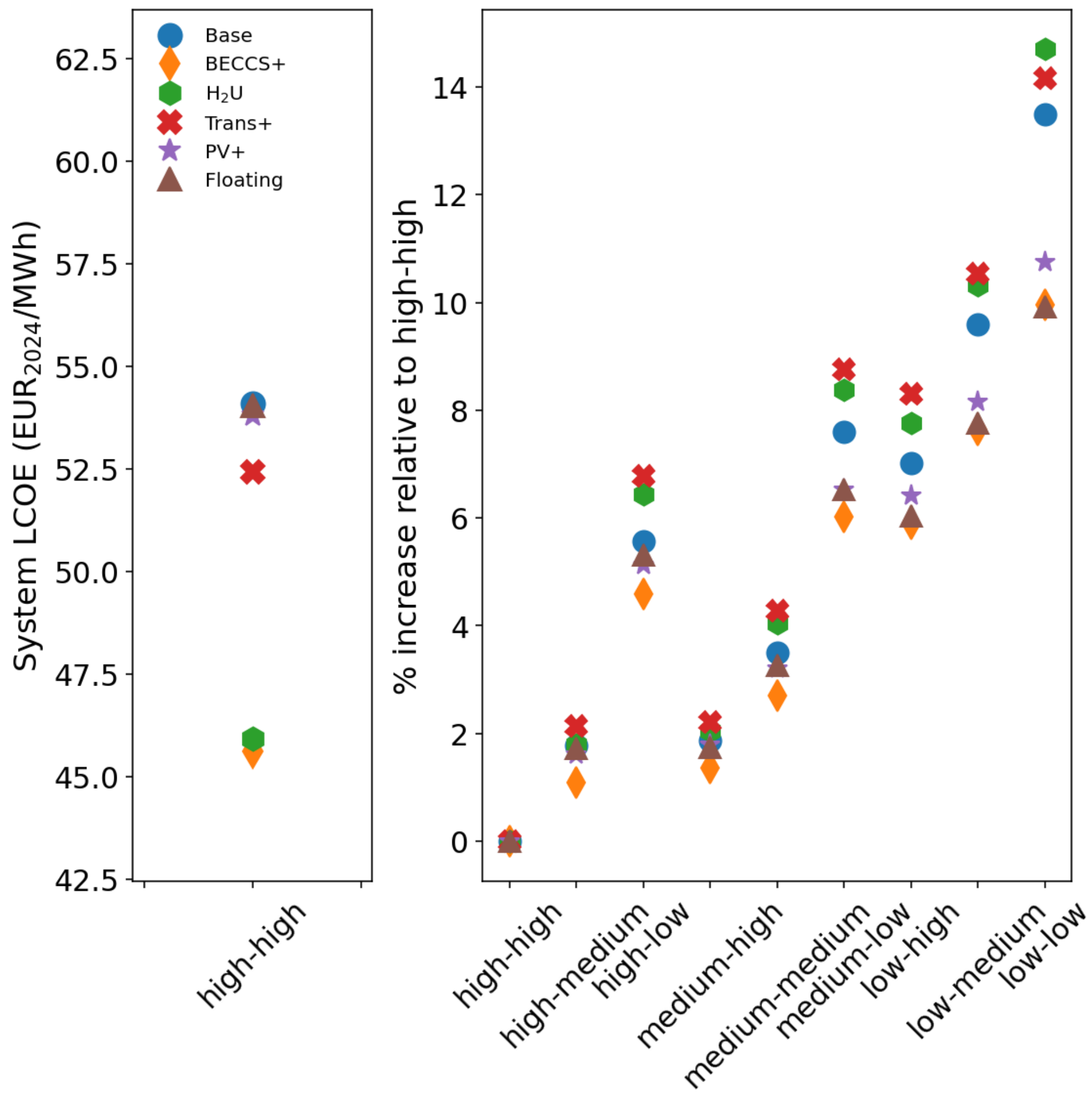

**Figure S2.1. Total European system levelised cost of electricity (left panel) and its increase across our impact tolerance scenarios (right panel) for the modelled sensitivity cases.**

# Supplementary note 3 – Methodological details of land availability scenarios

## Narrative and implementation

The scenario levels are based on a range of possible futures for wind energy deployment and the degree of tolerance to its social and environmental impacts by local communities. A low tolerance to the impacts of wind energy indicates a high contestation around new projects and therefore its overall deployment potential is very limited. There are examples from the recent past across Europe of situations where this has happened.
For example, in Norway, a rapid scale-up and deployment of onshore wind between 2017-2021 as well as a government issued national framework for wind power contributed to a nation-wide discussion on the impacts and perceived harm of wind power. The topic engaged large parts of society and was an active topic on both social media[5] and in newspapers[6] and lead to a previously unseen number of responses to a public consultation process[7]. This in turn led to a shift in political support for onshore wind where most political parties changed their stance and even a temporary halt in new licences until the framework had been re-designed. Norway went from deploying almost 1.3 GW (+61%) in 2020 and 1.1 GW in 2021, to 0.3 GW in 2022, 10 MW in 2024 and nothing in 2025[8].

In Austria, the 4.3 GW of existing wind power capacity is concentrated in three of the country's nine federal states. These are all located in the East, where installation is more straightforward due to flat terrain and landscape being considered less scenic compared to states in the Alps. As these low-hanging fruits have been exhausted, the permitting of new wind power installations has slowed down. One federal state with only 28 MW of wind power deployment (Carinthia in the South) held a referendum in 2025 in which 51.5% of voters rejected wind energy development. The referendum was initiated by the right-wing populist party FPÖ and opposed by industry, church, the climate advisory board and climate activist groups. Although the referendum is not legally binding, it has nonetheless generated strong political resistance but is seen not in line with constitutional law by constitutional judges.

The UK saw onshore wind power expand during the 2000s and early 2010s driven by key policies such as the financial support schemes the Renewables Obligation (RO) and Contracts for Difference (CfD). However, planning policy was changed in 2015 under the then Conservative Government to better speak to the concerns of the party's rural base by shifting the final say on projects greater than 50 MW from national to local level. Indeed, Government also mandated that there be demonstrable local public support for projects to be approved[9]. Furthermore, access to RO was cut in 2016 and CfDs in 2019[10]. As these changes filtered through annual capacity growth fell from a peak of 1.8 GW in 2017 to just 574 MW in 2019 and as low as 76 MW in 2020, or 95% reduction relative to peak, and has averaged 500 MW a year since[11].

Denmark is seen as a European and even global pioneer in wind energy. Nationally, it saw significant onshore wind penetration from the early 1990s which grew steadily until the early to mid-2000s, enabled by policy support and a history of co-operative ownership[12]. Peak capacity additions reached 590 MW in 2000[13,14]. During the 2000s, the role of wind farm co-operatives began to diminish as large developers and investors stepped in to deliver the next generation of larger turbines. Combined with political changes which led to the cutting back of financial incentives and new projects becoming increasingly locally contested, onshore capacity growth stagnated[12]. It was subsequently somewhat rejuvenated in the early 2010s partly because of The Renewable Energy Act and its attempts to enhance local acceptance. However, in the recent past capacity addition have slumped to 101 MW in 2022 and 9 MW in 2023[13,14]. This can partially be explained by larger turbines and projects together with more capacity in total leading to greater visual impact and growing noise concerns[15]. But is also related to ownership models moving away from community co-operatives, which served to embed local communities as co-creators of the sector, and small projects toward large utility scale farms owned by large private actors[15]. As a result, onshore wind has become a contested topic even in the home of European wind power.
As such, if the impacts of wind energy are not carefully handled, Europe might see more situations where wind energy deployment is severely limited, potentially realising the future that the low tolerance scenarios represent.

**Table S3.1 Summary of social and environmental exclusion levels**

| Dimension | Exclusion | High | Medium | Low |
|---|---|---|---|---|
| **Environmental** | Protected areas - CDDA | Ia, Ib, II, III, IV | Country specific[b] | All categories + 2km buffer |
| | Protected areas - Natura2000 | No | Country specific[b] | Yes + 2km buffer |
| | Bird vulnerability and threatened bat pseudo richness | No | Country specific[b] | All countries. Pixels with highest 40% of vulnerable species counts. |
| | Peatlands | No | No | Yes |
| | Forests | No | Country specific[b] | Yes |
| **Social** | Settlement setback[a] | 200m | Current regulations[b] | 3000m |
| | Noise | 250m | 500m | 2000m |
| | Shadow flicker | 84m | 1250m | 2500m |
| | Scenicness | No | Country specific (highly scenic and/or scenic)[b] | Highly scenic and scenic for all countries |
| | Coastline | 6 nautical miles (11,112m) | 12 nautical miles (22,224m) | 12 nautical miles (22,224m) |

[a] Settlement distances are based on the following CORINE land cover categories: Continuous urban fabric (1.1.1), Discontinuous urban fabric (1.1.2), Industrial or commercial units (1.2.1), Construction sites (1.3.3), Green urban areas (1.4.1) and Sport and leisure facilities (1.4.2)
[b] See Supplementary Note 5 for details.

# Technical exclusions

In addition to the exclusions based on social and environmental impacts, we also consider more non-negotiable technical factors that prevent wind energy deployment in certain places. For example, we do not consider that wind turbines can be placed on a road or very close to it. A summary of these technical exclusions is shown in Table S3.2 and its geographical spread in Figure S3.1.

**Table S3.2 Technical exclusions**

| Category | Description | Buffer distance | Data source |
|---|---|---|---|
| Onshore only exclusions | | | |
| Hydrology | Exclude water bodies and rivers | 175m | EU-Hydro River Network Database[16] |
| Airports | Exclude airports | 6,000m | OSM[17] |
| Slope | Exclude areas with a slope > 15° | None | Copernicus DEM GLO-90[18] |
| Powerlines | Exclude powerlines | 175m | OSM[17] |
| Roads | Exclude roads | 175m | OSM[17] |
| Railways | Exclude railways | 175m | OSM[17] |
| Radars | Exclude radars | 7,000m | OSM[17] |
| Glaciers | Exclude glaciers | None | OSM[17] |
| Ports | Ports | None | LUISA[19] |
| Both onshore and offshore | | | |
| Military areas | Exclude military areas | 5,000m | OSM[17] |
| Offshore only exclusions | | | |
| O&G platforms | Exclude oil and gas platforms | 2,000m | Martins et al.[20,21] |
| O&G pipelines | Exclude oil and gas pipelines | 150m | Martins et al.[20,21] |
| Shipping lanes | Exclude medium and heavily trafficked shipping lanes | None | eMODNET[22] |

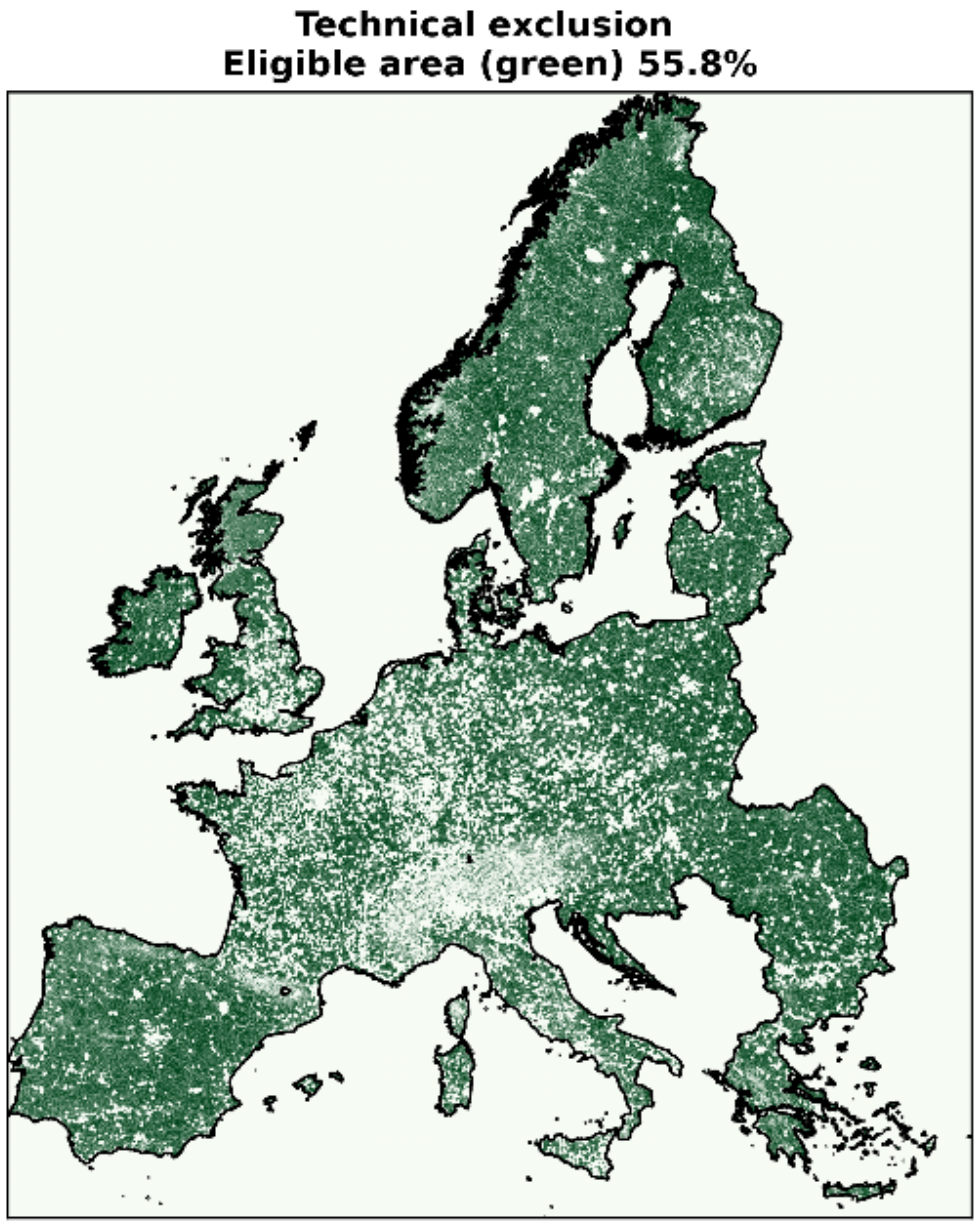

| Country | area ($km^2$) | share (%) |
|---|---|---|
| AT | 34940.4 | 41.6 |
| BE | 13075.0 | 42.6 |
| BG | 66856.4 | 60.2 |
| CH | 8264.9 | 20.0 |
| CZ | 36724.9 | 46.6 |
| DE | 163383.5 | 45.7 |
| DK | 21387.8 | 49.5 |
| EE | 28563.5 | 63.0 |
| ES | 314355.2 | 63.1 |
| FI | 222584.4 | 65.9 |
| FR | 238552.2 | 43.5 |
| GR | 73021.6 | 55.4 |
| HR | 35264.7 | 62.4 |
| HU | 56315.0 | 60.5 |
| IE | 48460.1 | 69.3 |
| IT | 110309.2 | 36.7 |
| LT | 41922.9 | 64.6 |
| LU | 1163.6 | 44.8 |
| LV | 41795.0 | 64.7 |
| NL | 15520.7 | 41.5 |
| NO | 196296.3 | 60.7 |
| PL | 199251.0 | 63.9 |
| PT | 53961.9 | 60.8 |
| RO | 168269.8 | 70.6 |
| SE | 287859.2 | 64.0 |
| SI | 10657.6 | 52.6 |
| SK | 27609.0 | 56.3 |
| UK | 120409.1 | 49.2 |

**Figure S3.1 Technical exclusions across the modelled European countries.**

# Supplementary note 4 – Methodological details of the modelling framework

## JRC-EU-TIMES policy constraints

JRC-EU-TIMES includes comprehensive representation of European legislation and policy instruments aimed to ensure net-zero greenhouse gas (GHG) emissions by 2050. This includes GHG limits, renewable and efficiency targets as well as sector-specific constraints.

First, JRC-EU-TIMES translates binding European legislation into internally consistent transition pathways. The model implements EU legislation in force as of 1 January 2024, including the European Climate Law (Regulation (EU) 2021/1119), the revised EU Emissions Trading System Directive (EU 2023/959), the Effort Sharing Regulation (EU 2023/857), the Renewable Energy Directive (EU 2023/2413), the Energy Efficiency Directive (EU 2023/1791), the Energy Performance of Buildings Directive (EU 2018/844), $CO_2$ standards for light- and heavy-duty vehicles (EU 2019/631; EU 2023/851), aviation-related provisions under the revised ETS and sustainable aviation fuel mandates, as well as nationally legislated coal phase-out commitments.  These instruments are operationalised through binding GHG caps, renewable and efficiency targets, and sector-specific constraints consistent with the Fit-for-55 package.

Additional constraints are imposed to ensure compliance with EU net-zero GHG emissions by 2050, enforcing at least a 55% net GHG reduction by 2030 and 90% by 2040 (relative to 1990). These limits include the treatment of Land use, land-use change, and forestry (LULUCF), and partial coverage of international transport as defined in the Climate Law. The revised EU-ETS cap trajectory and strengthened renewable and efficiency targets are applied accordingly. The legislation and policy instruments are summarised in Table S4.1

**Table S4.1 Overview of legislation and policy instruments included in JRC-EU-TIMES**

| Policy / Legal Act | Type of instrument | Key quantitative targets / provisions |
|---|---|---|
| **European Climate Law** (Regulation (EU) 2021/1119) | Economy-wide GHG cap and trajectory | Net-zero GHG by 2050; ≥ 55% net GHG reduction by 2030 vs 1990 |
| **Revised EU ETS Directive** (Regulation (EU) 2023/959) | Cap-and-trade | Strengthened ETS cap trajectory to 2030 with higher linear reduction factor; broadened sector coverage |
| **Effort Sharing Regulation** (Regulation (EU) 2023/857) | Non-ETS emission caps | More stringent 2030 reduction targets for non-ETS sectors, replacing the EU-wide -30% goal with differentiated national targets |
| **Renewable Energy Directive** (Directive (EU) 2023/2413) | Renewable share and sub-targets | At least 42.5% share of renewables in EU gross final energy consumption by 2030; minimum shares for advanced biofuels and; targets for renewable hydrogen in industry; annual increases in RES share for heating/cooling. |
| **Energy Efficiency Directive** (Directive (EU) 2023/1791) | Energy-efficiency / final-energy cap and savings obligation | Collective binding EU target: at least 11.7% reduction in final energy consumption by 2030 vs reference (maximum 992.5 Mtoe); annual energy-savings obligation of 1.49% (2024–2030); public bodies to reduce energy use by 1.9% per year vs 2021. |
| **Energy Performance of Buildings Directive** (Directive (EU) 2018/844) | Performance standard / technology requirement | From 2027: all new public buildings must be zero-emission; from 2030: all new buildings must be zero-emission; long-term renovation strategies and building renovation passports to improve building stock performance. |
| **$CO_2$ standards for light- and heavy-duty vehicles** (Regulation (EU) 2019/631; (EU) 2023/851) | Performance standard / technology mandate | -55% fleet-average $CO_2$ emissions for new cars and -50% for vans in 2030-2034 vs 2021; zero-emission vehicles from 2035. Up to -90% $CO_2$ reduction by 2040 vs 2019-2020; all new city buses to be zero-emission by 2030 |
| **Aviation-related provisions** | Cap-and-trade plus fuel-blending mandate | Aviation emissions increasingly covered by the ETS cap; minimum shares of sustainable aviation fuel in aviation fuel from 2025, with rising mandated shares towards 2050. |
| **National coal phase-out commitments** | Technology/phase-out constraint | Country-specific coal phase-out years, depending on Member State legislation. |

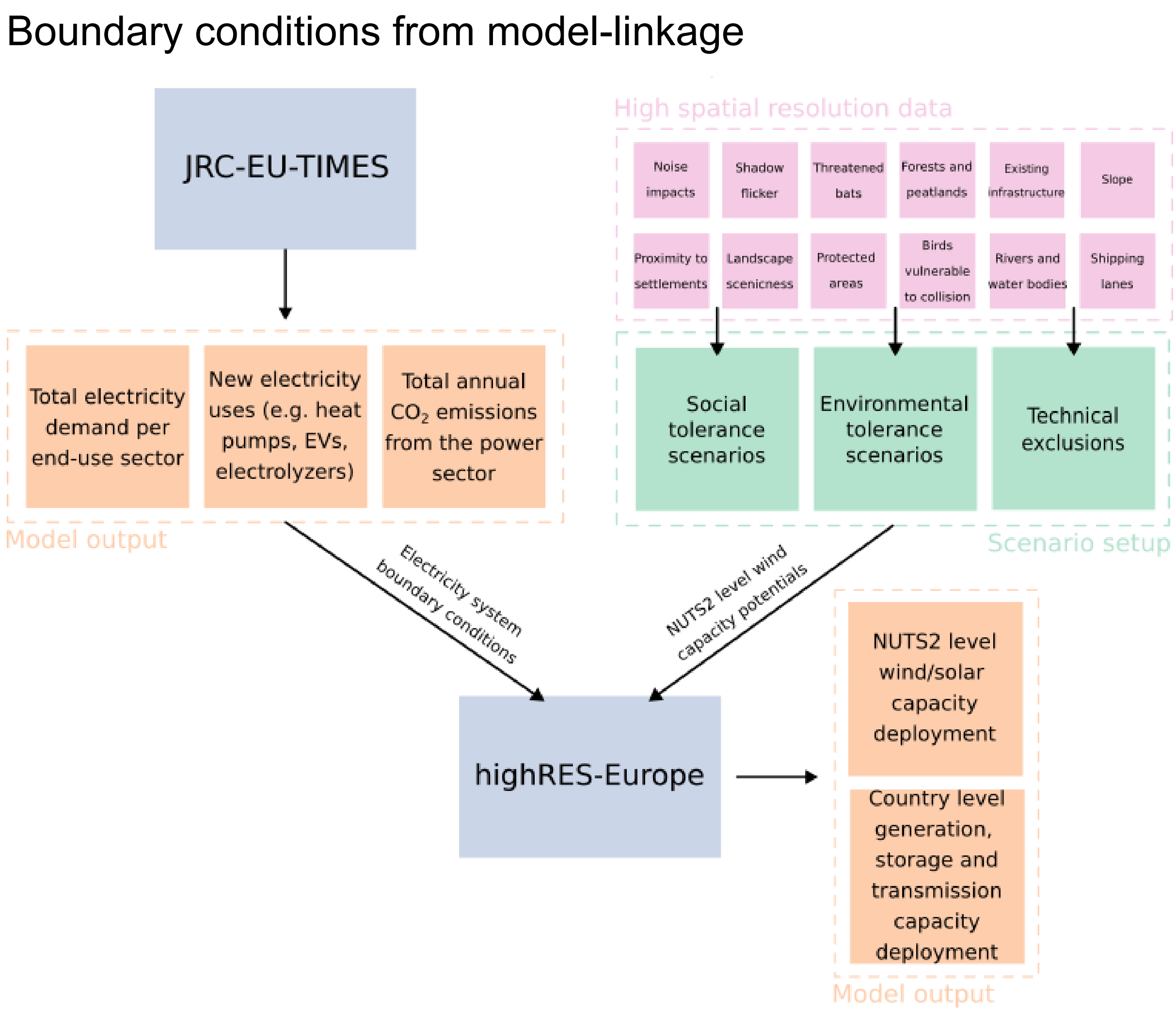

**Figure S4.1 Schematic overview of the modelling framework and the link from JRC-EU-TIMES to highRES-Europe**

**Table S4.2. Annual electricity demand and emissions budget for all EU25+3 countries in 2050 as modelled by JRC-EU-TIMES.**

| Country | Annual electricity demand in 2050 (TWh) | Annual $CO_2$ emissions budget for the electricity sector (kt) |
|---|---|---|
| AT | 143 | -45 |
| BE | 175 | -666 |
| BG | 62 | -5,153 |
| CH | 81 | -2,833 |
| CZ | 133 | -3,538 |
| DE | 1,128 | -5,160 |
| DK | 132 | -3,445 |
| EE | 13 | -772 |
| ES | 687 | -19,221 |
| FI | 186 | -4 |
| FR | 811 | -2,376 |
| GR | 134 | -3,279 |
| HR | 35 | -825 |
| HU | 91 | 0 |
| IE | 61 | -2,871 |
| IT | 668 | -6,670 |
| LT | 23 | -1,175 |
| LU | 11 | -269 |
| LV | 21 | -564 |
| NL | 461 | -2,505 |
| NO | 356 | 0 |
| PL | 359 | -7,201 |
| PT | 88 | -1,285 |
| RO | 117 | -2,508 |
| SE | 241 | -7,879 |
| SI | 24 | -300 |
| SK | 52 | -1,348 |
| UK | 773 | 627 |
| EU25+3 | 7,066 | -81,265 |

Hourly future electricity demand profiles for each country modelled in highRES-Europe are developed based on separate sectoral loads making up the total annual demand in Table S4.2, which are then added to historical demand from 2012. For electrified heat in buildings the approach follows that in Ref.[23], i.e. national level annual demands are downscaled using heating degree days based on ERA5 data across the year (1995) and assumed hourly air source heat pump load profiles over the day. Electric vehicles are also treated as they were in that study with national annual demands converted to the hourly level using assumed charging profiles. Further electrification of residential and commercial loads are modelled by multiplicatively scaling up historical total annual demand (i.e. before the addition of significant heat pump and electric vehicle penetration) by the required amount. Finally, increases in industrial electrification are represented by the simple approach of adding on a flat load profile which sums up to the required amount across the year for each nation.

# highRES-Europe technology assumptions

**Table S4.3 Technology options in the Base scenario in this study**

| Scenario | Type | Investable technologies available | Technologies with fixed installed capacities |
|---|---|---|---|
| Base | Generation | Nuclear<br>Natural gas CCGT with CCS<br>Natural gas OCGT<br>Biomass<br>BECCS<br>Solar PV<br>On/offshore wind | Hydropower reservoir<br>Hydropower run-of-river |
| | Storage | Li-ion batteries<br>Hydrogen storage in pressurised tanks | Pumped-hydro |

**Table S4.4. Technology cost parameters for highRES-Europe in 2024€.**

| Technology | Overnight CAPEX [€k/MW] | Discount rate [%] | Economic lifetime [years] | FOM [€k/MW-year] | VOM [€/MWh] |
|---|---|---|---|---|---|
| Onshore wind | 1320 | 5.2 | 25 | 35.4 | 9 |
| Offshore wind – Bottom fixed[1] | 1668[a] | 6.3 | 25 | 151 | 3 |
| Offshore wind - Floating[1] | 2197[a] | 7.15 | 25 | 151 | 3 |
| Solar PV | 430.5 | 5.0 | 25 | 8.86 | |
| Nuclear | 6957 | 9.0 | 60 | 112 | 8 |
| Biomass | 4535 | 7.15 | 25 | 185 | 8 |
| Biomass with CCS | 5060 | 7.15 | 30 | 185 | 7 |
| OCGT | 547.4 | 7.15 | 25 | 24.8 | 9 |
| CCGT with CCS | 2090 | 7.15 | 30 | 58.5 | 7 |
| Hydropower – Run-of-river | N/A | N/A | N/A | 74 | 4 |
| Hydropower - Reservoir | N/A | N/A | N/A | 74 | 4 |
| Li-ion batteries (Power) | 101.0 | 7.15 | 13 | 12.4 | |
| Li-ion batteries (Energy) | 118.7 (€k/MWh) | 7.15 | 13 | | |
| H2 surface tanks or underground storage (Power: Electrolyser + H2OCGT) | 777.7 | 7.15 | 25 | 24.8 | 9 |
| H2 surface tanks or underground (Power: Electrolyser + H2CCGT) | 1224 | 7.15 | 25 | 53.1 | 7 |
| H2 surface tanks | 13.73[b] (€k/MWh) | 7.15 | 25 | | |
| H2 salt cavern | 0.4783[c] (€k/MWh) | 7.15 | 40 | | |
| HVAC overhead interconnector | 80.13[d] (€k/MW-100km) | 7.15 | 40 | 2% of capex | |
| HVAC substation | 198[e] | 7.15 | 40 | | |
| HVDC marine interconnector | 176[e] (€k/MW-100km) | 7.15 | 40 | 2% of capex | |
| HVDC substation | 523.9[e] | 7.15 | 40 | | |

[1] There is an additional grid connection with HVDC line cost as above and offshore platform/converter and onshore converter of 838[e] €k/MW.

Data sources are as in Ref.[23] apart from:

[a] Danish Energy Agency - Version 16 - Data sheet for electricity and district heating production[24]
[b] Ref.[25]
[c] Ref.[26]
[d] Ref.[27]
[e] Ref.[28]

# highRES-Europe spatial topology

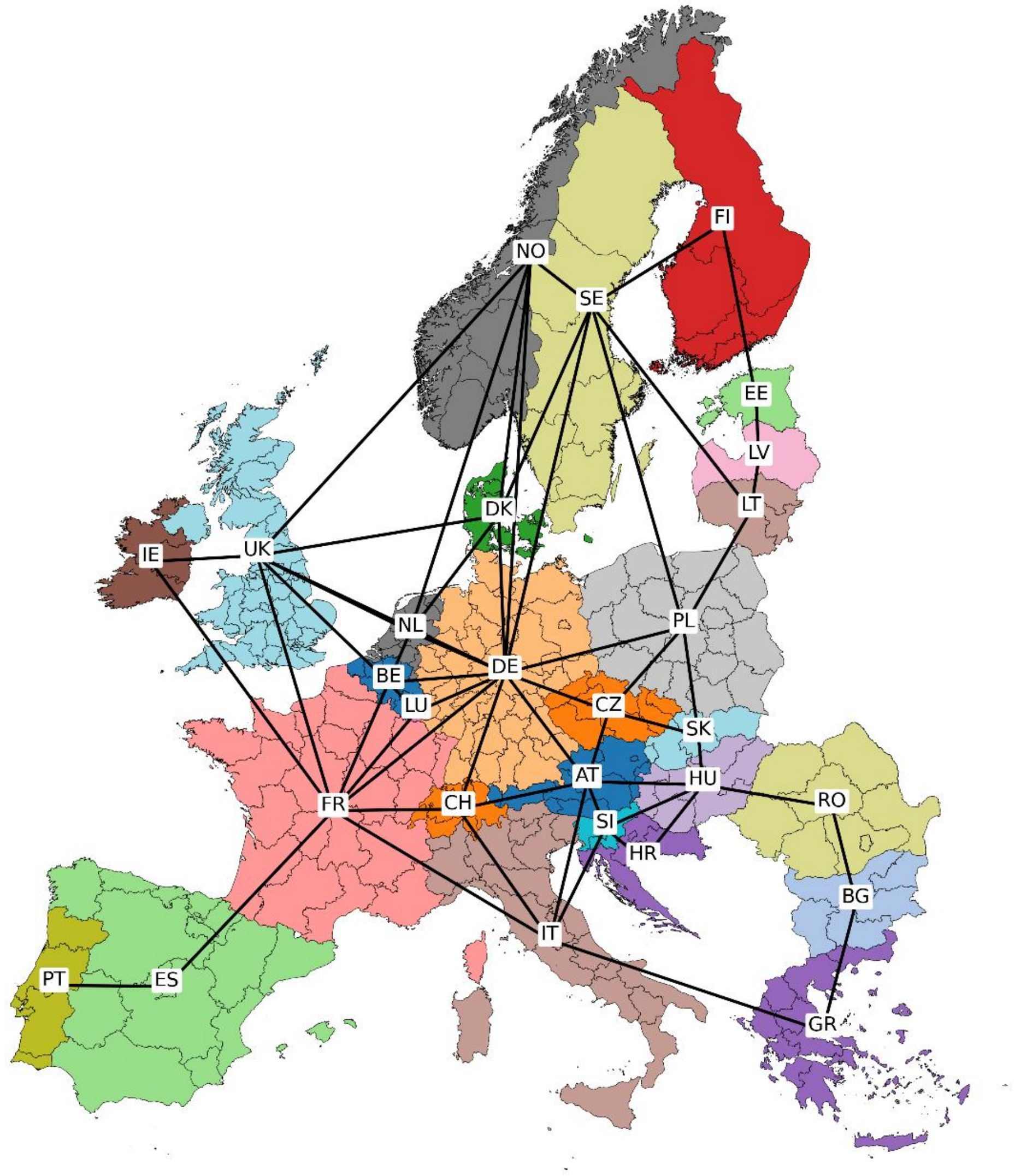

**Figure S4.2 highRES-Europe spatial topology.** Map showing the transmission network that connects the EU25+3 countries modelled in this work. Each country is also shown broken out into its NUTS2 regions, the spatial resolution at which onshore/offshore wind and solar PV capacity deployment is modelled here.

# Bias-correction

highRES-Europe uses ERA5 reanalysis data (0.25°×0.25°)[29] to compute wind speed at turbine hub height, calculate capacity factors at the grid-cell level, and then aggregate them to the corresponding NUTS2 regions. Even though this dataset provides a good representation of wind speed, it also exhibits a well-known bias in areas with complex topography, such as the Alps and the Scandinavian Mountains[30]. To mitigate this issue, we use bias-correction ratios[31], computed by dividing microscale wind speed values from the Global Wind Atlas (GWA)[32] on a 0.025ºx0.025º grid by the corresponding ERA5 wind speed values averaged over 20 years (2008-2017), following methodology described in Ref.[33,34]. By applying these ratios to ERA5 100m wind speeds in a given year, we then correct and downscale the data to 0.025°×0.025° resolution. To ensure consistency with other ERA5 weather variables used in highRES-Europe, such as solar irradiance, we resample the resulting values back to the original ERA5 resolution.

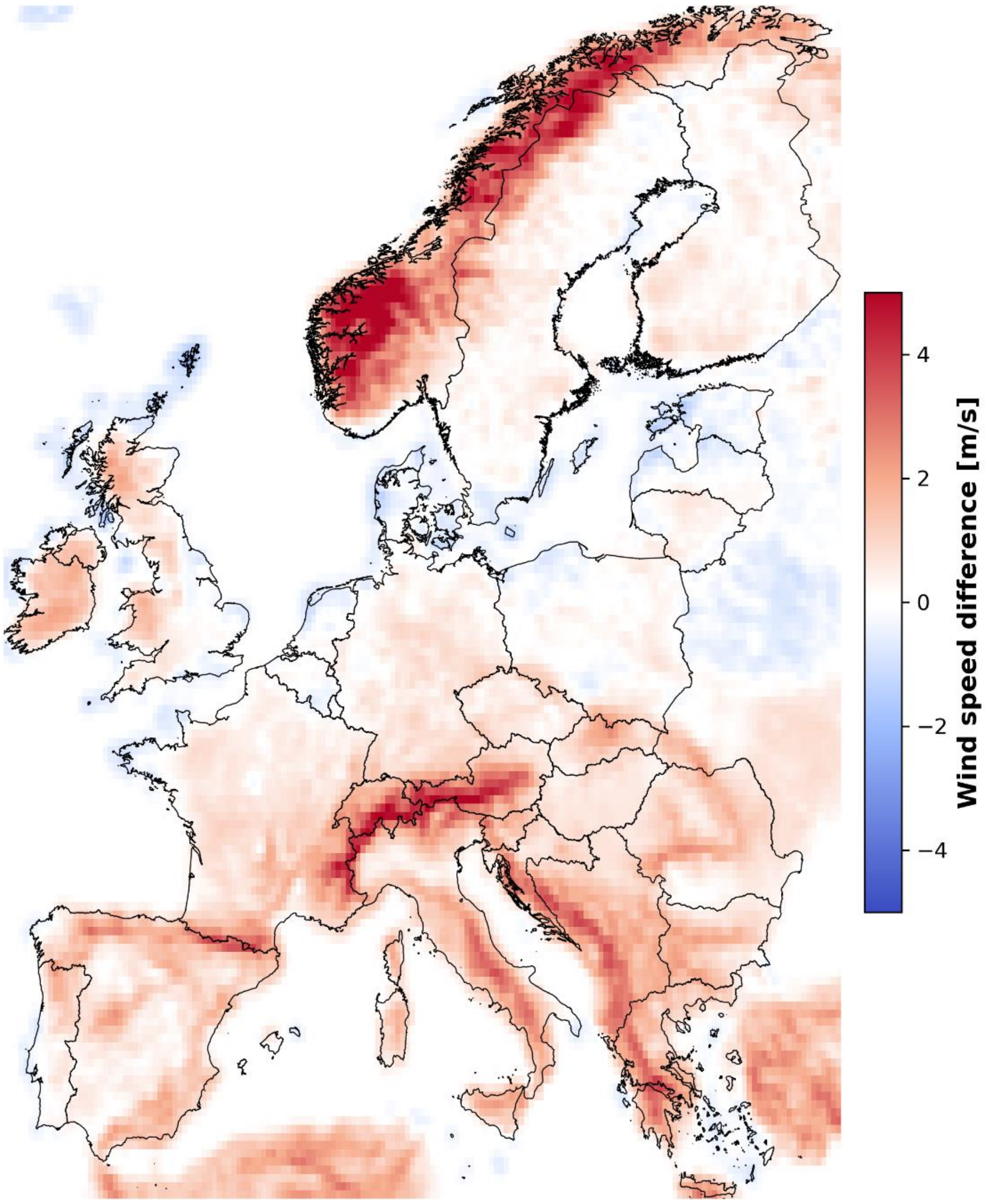

**Figure S4.3: Difference between the mean ERA5 wind speed (m/s) corrected by GWA ratios and the original ERA5 data at 0.25ºx0.25º resolution for the year 1995.** The red colour highlights areas where the wind speed from ERA5 is underestimated, and the blue colour indicates where it is overestimated.

Figure S4.3 illustrates the difference between bias-corrected 100m wind speed, obtained using GWA-derived ratios, and the original ERA5 data. Grid cells coloured red show areas where ERA5 underestimates wind speed, while cells coloured blue indicate overestimations. As noted earlier, ERA5 generally underestimates wind speed across much of continental Europe, particularly in Norway and the Alps, but it likely overestimates wind speed in the coastlines of the UK.

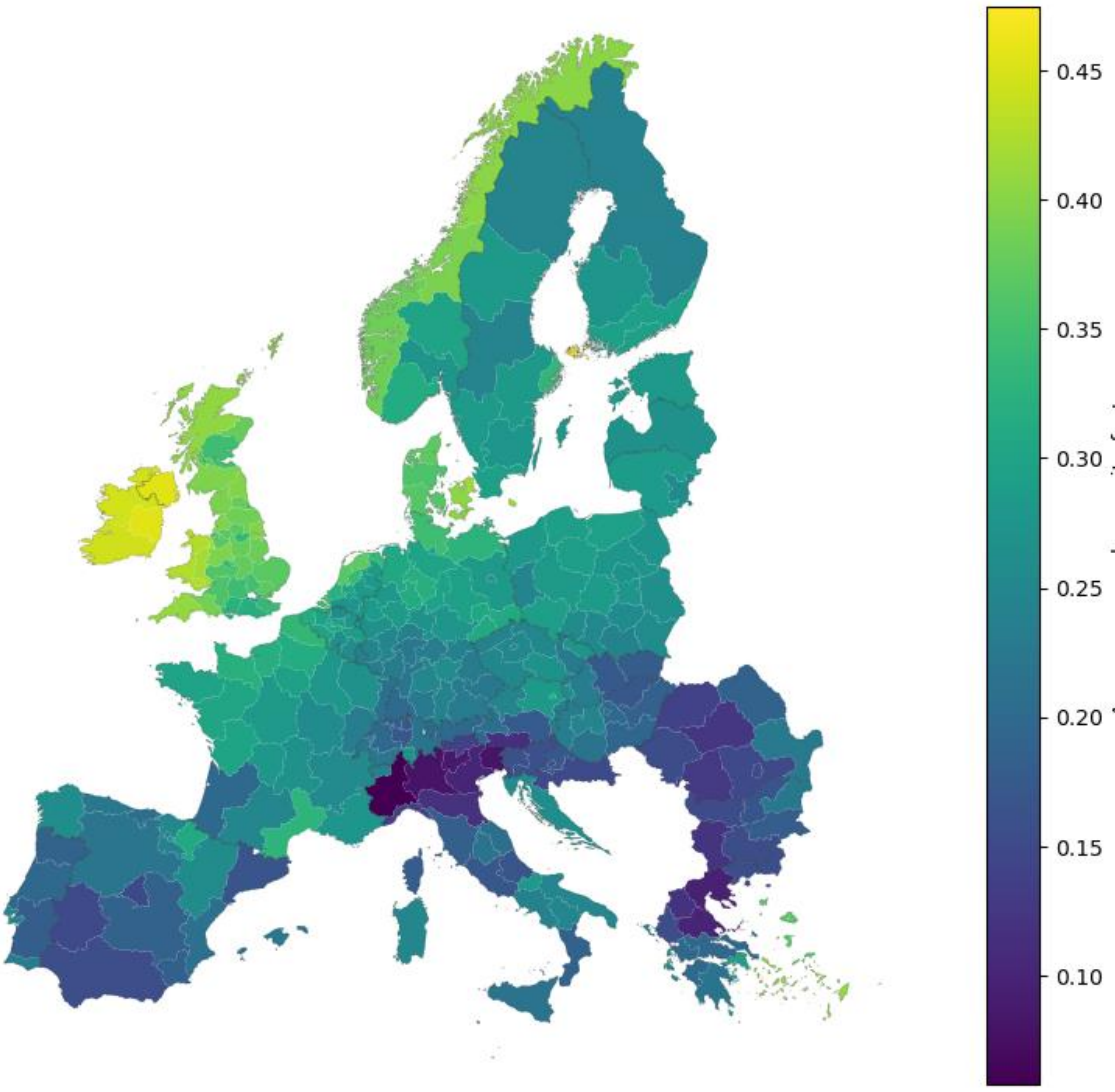

**Figure S4.4. Average annual capacity factors in the H-H scenario.**

The resulting average capacity factors for each NUTS2 region after bias-correction and land-availability exclusions are shown in Fig S4.4.

# Solar PV growth rate constraints

The deployment of VREs is not only limited by the land available for the technology, but also by factors such as material scarcity, grid integration or financial profitability[35]. Whereas common approaches to technology growth rates include assuming a logistic growth (commonly referred to as an S-curve, or an asymmetric S-curve, a Gompertz curve)[36], the aim of our restrictions is to approximate simple and plausible upper limits for deployment in individual countries. A recent study by Vetier et al.[35] on historic deployment of onshore wind in large European countries has suggested a modification of the S-curve, where onshore wind deployment follows growth pulses of typically four to eight years. Furthermore, the study found that "[...] the growth rates of individual pulses rarely exceeded 1.5 times the overall maximum growth rate derived from a single-fit model for the whole curve.". Although the cumulative deployments of 25 years of high technology deployment may be overly optimistic, our assumption of 1.5 times the peak growth rate extrapolated to 2050 is only meant as an upper bound.

For Finland and Norway, solar PV still contributes less than one percent of the total electricity generation, meaning that they are still in their "formative phase" with erratic growth patterns[37]. We account for this by replacing both Norway and Finland's maximum 2050 capacity with neighbouring country Sweden. Sweden, Finland and Norway are situated at similar latitude and have about the same population and land-area.

**Table S4.5 Installed capacity of solar PV, peak growth rate and resulting maximum capacity in 2050. Data based on IRENA[1].**

| Country | Technology | 2024 capacity [GW] | Peak growth rate [GW/a] (year) | Maximum 2050 capacity [GW] |
|---|---|---|---|---|
| Austria | Solar PV | 8.5 | 2.6 (2023) | 108.9 |
| Belgium | Solar PV | 9.8 | 1.6 (2023) | 71.0 |
| Bulgaria | Solar PV | 3.9 | 1.2 (2023) | 49.6 |
| Czechia | Solar PV | 4.2 | 1.3 (2010) | 53.4 |
| Croatia | Solar PV | 0.9 | 0.4 (2024) | 16.3 |
| Denmark | Solar PV | 3.9 | 1.4 (2022) | 57.2 |
| Estonia | Solar PV | 1.3 | 0.5 (2024) | 21.3 |
| Finland | Solar PV (mod) | 1.2 | 1.6 | 63.8 |
| | Solar PV | 1.2 | 0.3 (2023) | 14.4 |
| France | Solar PV | 21.5 | 4.1 (2024) | 182.6 |
| Germany | Solar PV | 89.9 | 15.1 (2024) | 677.3 |
| Greece | Solar PV | 9.3 | 2.6 (2024) | 109.9 |
| Hungary | Solar PV | 7.7 | 1.8 (2024) | 77.5 |
| Ireland | Solar PV | 1.3 | 0.5 (2023) | 22.5 |
| Italy | Solar PV | 36.0 | 9.4 (2011) | 404.4 |
| Latvia | Solar PV | 0.5 | 0.2 (2023) | 8.5 |
| Lithuania | Solar PV | 2.6 | 1.2 (2024) | 51.0 |
| Luxembourg | Solar PV | 0.5 | 0.1 (2024) | 5.6 |
| Norway | Solar PV (mod) | 0.8 | 1.6 | 63.4 |
| | Solar PV | 0.8 | 0.3 (2023) | 13.0 |
| Netherlands | Solar PV | 24.0 | 3.9 (2023) | 176.9 |
| Poland | Solar PV | 20.2 | 5.4 (2023) | 229.4 |
| Portugal | Solar PV | 5.8 | 1.8 (2024) | 74.8 |
| Romania | Solar PV | 4.7 | 1.7 (2024) | 71.0 |
| Spain | Solar PV | 36.3 | 6.1 (2023) | 272.0 |
| Slovenia | Solar PV | 1.3 | 0.4 (2023) | 17.1 |
| Slovakia | Solar PV | 0.9 | 0.5 (2011) | 19.5 |
| Sweden | Solar PV | 5.0 | 1.6 (2023) | 67.6 |
| Switzerland | Solar PV | 7.8 | 1.7 (2024) | 74.1 |
| United Kingdom | Solar PV | 17.6 | 4.1 (2015) | 176.5 |
| EU25+3 | Solar PV (mod) | 327.3 | | 3,222.8 |
| | Solar PV | 327.3 | | 3,123.0 |

Tablenote: The row for EU25+3 is only a summary of the full spatial extent and no limit or growth rate is considered for the European system as a whole.

## Existing capacity

We update the existing capacity infrastructure for renewable energy at the NUTS2 level and for nuclear energy at the country level to more accurately reflect the electricity system in 2050. This improvement enables highRES-Europe to distinguish between existing and new capacity when making investment decisions, resulting in a system configuration that better aligns with projected future capacity.

We estimate existing capacity using data from the Global Energy Monitor (GEM)[38–40], an open-access platform that provides detailed information on worldwide power generation facilities. This database includes onshore wind, offshore wind, solar, and nuclear projects, with details such as project names, operational status, locations (latitude and longitude), and commissioning and decommissioning dates.

To identify the existing facilities in 2050, we begin by extracting power plant data for the 28 European countries included in the model (25 EU countries plus Norway, Switzerland, and the UK). We then project their future status based on commissioning dates and expected operational lifetimes. The commissioning date is directly retrieved from the GEM database. We assume a lifetime of 30 years for renewable technologies, wind and solar, and 40 years for nuclear plants, based on the 75th percentile of retired plant lifetimes from GEM. Using these parameters, we estimate retirement dates for plants currently operational or under construction by adding the expected lifetime to their commissioning dates. Finally, we exclude all plants expected to retire before 2050.

To aggregate the estimated infrastructure data at the NUTS2 level, each plant is assigned to its corresponding NUTS2 region by overlaying the plant coordinates in GEM with the correct region. For onshore plants, we use the NUTS2 region that matches their exact location. For offshore plants, we assign them to the nearest coastal NUTS2 region as the reference area. Using this classification, capacities are aggregated by NUTS2 region, except for nuclear capacities, which are summed at the country level.

Figure S4.5 shows the spatial distribution of current capacity expected to be operational in 2050. Onshore wind capacity is distributed across Europe, with a higher concentration in Nordic countries. Solar capacity mainly exists in southern regions, with Spain having the highest concentration. Offshore wind capacities are primarily located in the North Sea area. Nuclear capacities are limited and are found in Finland, France, Slovakia, and the UK.

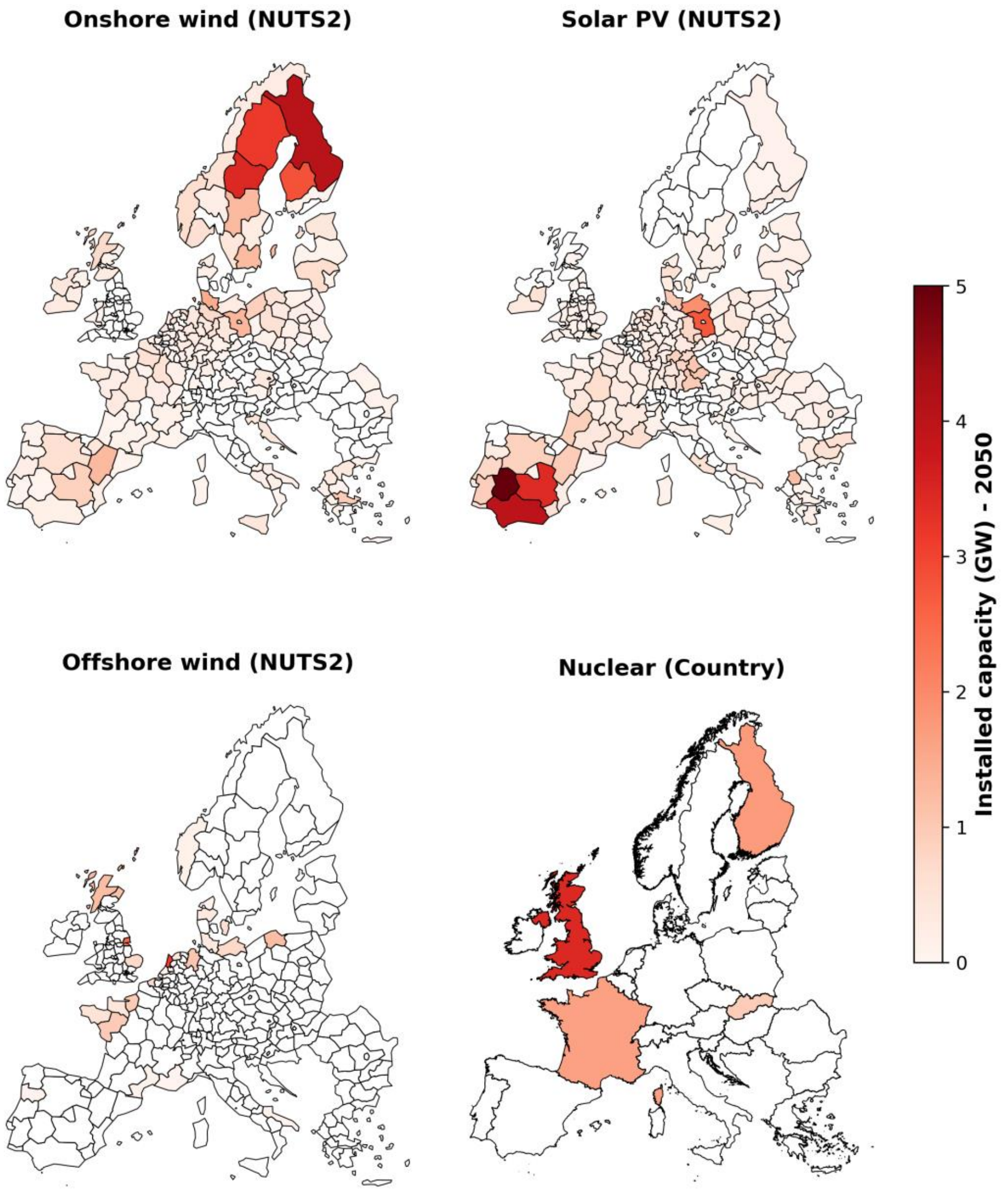

**Figure S4.5. Estimated existing capacity for onshore and offshore wind, solar PV, and nuclear in 2050.**

## Onshore wind land area to capacity potential

One of the goals of highRES-Europe is to determine, based on the available area in each NUTS2 region, where new renewable energy capacity is located. To do this, highRES-Europe receives the capacity potential per NUTS2, derived from land availability constraints, along with a conversion factor (capacity density) for each technology across Europe. In this work, onshore wind has a capacity density of 2.4 MW/km$^2$ using wind farm footprints and capacities reported in ref.[41] and following the next steps. First, we calculate the capacity density for each wind farm by dividing its capacity by its footprint. Second, we aggregate the data at the country level by taking the median across wind farms within each country. Third, we compute a capacity density-weighted average across countries using the normalised number of wind farms in each country as weights (dividing the number of wind farms in each country by the total number of wind farms at the continental level). Table S4.6 shows the country-aggregated values for this factor and the resulting weighted average used in this work.

**Table S4.6 Country-specific and European capacity density values for onshore wind based on the wind farm footprints reported in Ref.[41].**

| Country | Capacity density [MW/km$^2$] | Num. of WFs |
|---|---|---|
| *Austria* | 2.46 | 41 |
| *Belgium* | 2.35 | 75 |
| *Bulgaria* | 1.07 | 24 |
| *Croatia* | 0.96 | 29 |
| *Czech Republic* | 4.12 | 47 |
| *Denmark* | 4.96 | 41 |
| *Estonia* | 3.4 | 3 |
| *Finland* | 1.46 | 107 |
| *France* | 2.4 | 501 |
| *Germany* | 3.39 | 480 |
| *Greece* | 1.94 | 41 |
| *Hungary* | 0.72 | 10 |
| *Ireland* | 1.89 | 47 |
| *Italy* | 1.29 | 22 |
| *Latvia* | 0.79 | 1 |
| *Lithuania* | 1.69 | 18 |
| *Luxembourg* | 3 | 3 |
| *Netherlands* | 2.16 | 77 |
| *Norway* | 1.18 | 52 |
| *Poland* | 1.13 | 114 |
| *Portugal* | 3.73 | 14 |
| *Romania* | 1.62 | 28 |
| *Slovakia* | 8.25 | 1 |
| *Slovenia* | 45 | 1 |
| *Spain* | 1.1 | 88 |
| *Sweden* | 1.28 | 117 |
| *Switzerland* | 10 | 15 |
| *United Kingdom* | 1.51 | 61 |
| ***Europe*** | **2.4** | 2058 |

# Supplementary note 5 – Country-specific exclusions and the construction of the medium tolerance scenarios

In the medium tolerance scenarios, we align social (setback distances to residential areas, noise, shadow flicker, and landscape visual impact) and environmental (protected areas, forests, and vulnerability of bird species to collision and threatened bat species distributions) considerations for onshore wind deployment in Europe with the aggregation of past practices for siting wind farms, and country-specific regulations and guidelines, with the goal of representing the planning decisions to date in each considered country. We follow the national policy and existing literature on noise, shadow flicker, and setback distances from settlements. For landscape visual impact, bird and bat vulnerability, protected areas, and forests categories, we conduct an overlap analysis of onshore wind infrastructure spatial footprints over the past 20 years and high-resolution data for Europe. In the following subsections, we describe the methodology and dataset used, as well as the resulting exclusion assumptions for social and environmental considerations.

A summary of the country-specific exclusion restrictions for every category and country is shown in Table S5.7.

# Existing regulations and guidelines for wind energy deployment

National regulations and guidelines on wind energy siting can vary significantly across countries (and even within the same country), including setback distances to residential areas, noise limits, and visual impact mitigation measures, to address local acceptance, and environmental and health concerns[42]. To represent this variation, we align the exclusion restrictions for the medium scenarios with the regulations and existing literature values for setback distances, noise, and shadow flicker buffers around settlements. To identify settlements in Europe, we use the CORINE land-cover type database[43], specifically categories Continuous urban fabric (1.1.1), Discontinuous urban fabric (1.1.2), Industrial or commercial units (1.2.1), Construction sites (1.3.3), Green Urban areas (1.4.1), and Sport and leisure facilities (1.4.2).

To capture the existing limits on noise and shadow flicker disturbance for onshore wind projects in Europe, we use reported values from the literature and European regulatory documents and set an exclusion buffer around settlements. For noise, the buffer is defined in accordance with Ref.[44], with a typical reported distance of approximately 500 m, corresponding to a noise level of 45-50 db(A)[44,45] for large wind turbines. For shadow flicker, we define a buffer of 1250 m based on model results from Ref.[46,47] and Ref.[48] to ensure a threshold exposure time of less than 30 hours per year.

To represent the variations among countries, particularly with respect to setback distances to residential areas, we utilise country-specific data on regulations and guidelines that wind farm developers follow when constructing new wind infrastructure in Europe, all of which are publicly available in Ref.[49]. More details in the data collection methodology and further analysis in Ref.[50]. This dataset reports regulations and guidelines for setback distances across countries and NUTS2 regions for various types of infrastructure, including buildings, airports, and military areas.

From this dataset, we exclude all categories except buildings, as well as some subcategories, such as “camping areas”, “allotments in the grassland”, “buildings worth preserving in the grassland”, “permanently managed shelters”, and “cultural goods world heritage”, and then calculate the mean value across each country. Table S5.1 presents country-specific regulations and guidelines aggregated by country on setback distances for onshore wind in Europe. As certain countries, such as Estonia and Luxembourg, are not included in this dataset, and some reported countries, such as Switzerland, Belgium, and the UK, have small values (below the 10th percentile), we use the average values from the reported countries for these cases.

**Table S5.1: Country-specific setback distances to residential areas. Values are based on regulations and guidelines reported in Ref.[50].**

| Country | Setback distance (m) |
|---|---|
| ***Austria*** | 1063 |
| ***Belgium*** | 628 |
| ***Bulgaria*** | 500 |
| ***Switzerland*** | 628 |
| ***Czechia*** | 500 |
| ***Germany*** | 897 |
| ***Denmark*** | 600 |
| ***Greece*** | 1000 |
| ***Spain*** | 625 |
| ***Finland*** | 950 |
| ***France*** | 500 |
| ***Croatia*** | 700 |
| ***Hungary*** | 700 |
| ***Ireland*** | 500 |
| ***Italy*** | 550 |
| ***Lithuania*** | 600 |
| ***Latvia*** | 800 |
| ***Netherlands*** | 400 |
| ***Norway*** | 700 |
| ***Poland*** | 700 |
| ***Portugal*** | 500 |
| ***Romania*** | 450 |
| ***Sweden*** | 750 |
| ***Slovenia*** | 500 |
| ***Slovakia*** | 1000 |
| ***United Kingdom*** | 628 |
| ***Estonia*** | 628 |
| ***Luxembourg*** | 628 |

# Wind farm overlap analysis

To examine how the aggregation of onshore wind farm planning decisions over the last 20 years has incorporated environmental and social considerations into their historical practices throughout Europe, we utilise a harmonised geospatial database of wind energy projects derived from an innovative satellite imagery methodology (refer to Ref.[41] for detailed information). This dataset provides comprehensive details on location, capacity, hub height, estimated construction time, decommissioned date, and NUTS2 territorial classification, among others, for wind farm projects across Europe, along with an estimate of their spatial footprint.

From this dataset, we focus this analysis on onshore wind infrastructure data from 2000 to 2023 and overlay their derived spatial footprint with high-resolution data on landscape aesthetics, protected areas, forests, and the vulnerability of bird and bat species. We aggregate, at the country level, the fractional overlap between existing wind farm footprints and each environmental and social consideration across the 28 European countries included in our model (25 EU countries, Norway, Switzerland, and the UK).

We then examine the distribution across countries and establish a threshold to determine whether, despite the absence of formal legislation preventing wind deployment, a category has been generally avoided in the country's planning decisions. If a country-specific value for that category falls below the threshold, it has been avoided in wind deployment practices and, therefore, excluded from the land availability calculation for the medium scenario. Conversely, if the value exceeds the threshold, the category has not been avoided in the planning decisions and, therefore, it is included in the land availability for that country. In this study, we use a 15% overlap threshold applied across all categories. This was derived as follows. Firstly, we compute the country-level overlap between the scenicness data (including “scenic” and “highly scenic” areas), Natura 2000 protected areas, forests, and the grid cells that contain the highest 25% pseudo-species richness for birds and bats. Then, separately for each of the five categories, we apply the Jenks natural break method with three clusters to segment the fractional overlap and identify the width of a cluster around zero. Finally, we derive the final threshold by averaging the first and second clusters across all five categories.

Additionally, for countries with fewer wind farms, like Slovakia and Slovenia, we assume a minimum number of wind farms based on the category for it be considered in the analysis. The next section provides a detailed description of the dataset and the findings of this analysis.

## Protected areas

To evaluate the siting decision of wind energy infrastructure over designated conservation zones, we overlap the wind farm footprint dataset with geospatial protected areas from Natura 2000[51] and from the Common Database on Designated Areas (CDDA)[52,53].

For Natura 2000, we aggregate all reported categories and then intersect them with the European wind farm fleet footprints, while for the CDDA, we examine each reported category separately (graded by the International Union for Conservation of Nature (IUCN)). Tables S5.2 and S5.3 show the country-aggregated fractional overlap for each country for Natura 2000 and CDDA, respectively.

At a 15% threshold and excluding countries with fewer than 5 wind farms, only Bulgaria has deployed wind projects within CDDA-protected areas (category: *notApplicable*). In this case, we include only the “notApplicable” areas within Bulgaria’s CDDA-protected areas, excluding the rest of the CDDA areas for wind deployment. In contrast, within Natura 2000 protected areas, countries such as Bulgaria, Croatia, and Portugal have some wind farm projects (overlap exceeding 15%). Therefore, we exclude Natura 2000 protected areas from almost all countries, except for those with more than 15% overlap.

## Table S5.2: Fractional overlap of wind farm footprint aggregated by country for Natura 2000.

| Country | Natura 2000 (%) | Num. of WFs |
|---|---|---|
| Austria | 3.4 | 85 |
| Belgium | 4.6 | 176 |
| Bulgaria | 38.9 | 49 |
| Croatia | 41.3 | 32 |
| Czechia | 9.8 | 48 |
| Denmark | 1.5 | 496 |
| Estonia | 0 | 3 |
| Finland | 0.2 | 141 |
| France | 4 | 1070 |
| Germany | 2.7 | 2383 |
| Greece | 29 | 142 |
| Hungary | 2.6 | 13 |
| Ireland | 17.9 | 162 |
| Italy | 7.4 | 259 |
| Latvia | 0.1 | 8 |
| Lithuania | 0 | 34 |
| Luxembourg | 11.2 | 10 |
| Netherlands | 5.8 | 178 |
| Norway | 0 | 64 |
| Poland | 1.4 | 320 |
| Portugal | 28.7 | 179 |
| Romania | 11 | 59 |
| Slovakia | 0 | 1 |
| Slovenia | 99.4 | 2 |
| Spain | 8.5 | 416 |
| Sweden | 1.5 | 721 |
| Switzerland | 0 | 18 |
| United Kingdom | 0.9 | 517 |

## Table S5.3: Fractional overlap of wind farm footprint aggregated by country CDDA.

| Country | Ia (%) | Ib (%) | II (%) | III (%) | IV (%) | V (%) | VI (%) | notReported (%) | notAssigned (%) | notApplicable (%) | Num. of WFs |
|---|---|---|---|---|---|---|---|---|---|---|---|
| Austria | 0 | 0 | 0 | 0 | 0 | 2.7 | 0 | 0 | 0 | 0 | 85 |
| Belgium | 0 | 0 | 0 | 0 | 1.2 | 0 | 1 | 0 | 0 | 0.5 | 176 |
| Bulgaria | 0.4 | 0 | 0 | 0 | 0 | 0.7 | 0.2 | 0 | 0 | 79.2 | 49 |
| Croatia | 0 | 0 | 0 | 0 | 0 | 0 | 0 | 0.5 | 0 | 0 | 32 |
| Czechia | 0 | 0 | 0 | 0 | 0.2 | 0.4 | 0 | 0 | 0 | 0 | 48 |
| Denmark | 0 | 0 | 0 | 0 | 1.1 | 0.1 | 0 | 0 | 0 | 12.5 | 496 |
| Estonia | 0 | 0 | 0 | 0 | 0 | 0 | 0 | 0 | 0 | 1.8 | 3 |
| Finland | 0 | 0 | 0 | 0 | 0.6 | 0 | 0 | 0 | 0.2 | 0 | 141 |
| France | 0 | 0 | 0 | 0 | 0.1 | 5.6 | 0 | 0 | 0 | 0 | 1070 |
| Germany | 0 | 0 | 0 | 0 | 0.5 | 11.2 | 0 | 0 | 0.7 | 0 | 2383 |
| Greece | 0 | 0 | 0 | 0 | 7.1 | 0 | 2.6 | 0 | 0 | 0 | 142 |
| Hungary | 0 | 0 | 0 | 0 | 0 | 0 | 0 | 0 | 0 | 0 | 13 |
| Ireland | 0 | 0 | 0.3 | 0 | 5 | 0 | 0 | 0 | 0 | 0 | 162 |
| Italy | 0 | 0 | 0.3 | 0 | 1.2 | 0 | 0 | 0 | 0 | 0 | 259 |
| Latvia | 0 | 0 | 0 | 0 | 0.1 | 0 | 0 | 0 | 0 | 0 | 8 |
| Lithuania | 0 | 0 | 0 | 0 | 0 | 0 | 0 | 0 | 0 | 0 | 34 |
| **Luxembourg** | 0 | 0 | 0 | 0 | 11.8 | 0 | 9.6 | 0 | 0 | 0 | 10 |
| **Netherlands** | 0 | 0 | 1.4 | 0 | 5.8 | 0 | 0 | 0 | 0 | 10.7 | 178 |
| **Norway** | 0.3 | 0 | 0 | 0 | 0 | 0 | 0 | 0 | 0 | 0 | 64 |
| **Poland** | 0 | 0 | 0 | 0 | 0 | 0 | 0 | 0 | 4.3 | 0 | 320 |
| **Portugal** | 0 | 0 | 0 | 0 | 0 | 5.9 | 0 | 0 | 0 | 0 | 179 |
| **Romania** | 0 | 0 | 0 | 0 | 0 | 0.9 | 0 | 0 | 0 | 0 | 59 |
| **Slovakia** | 0 | 0 | 0 | 0 | 0 | 100 | 0 | 0 | 0 | 0 | 1 |
| **Slovenia** | 0 | 0 | 99.4 | 75.8 | 0 | 0 | 0 | 0 | 0 | 99.4 | 2 |
| **Spain** | 0 | 0 | 0 | 0 | 0.3 | 0.6 | 0 | 0 | 0.3 | 0 | 416 |
| **Sweden** | 0.4 | 0.5 | 0 | 0 | 0.2 | 0.1 | 0 | 0 | 1 | 0.3 | 721 |
| **Switzerland** | 0 | 0 | 0 | 0 | 0 | 0 | 0 | 0 | 0.5 | 0 | 18 |
| **United Kingdom** | 0 | 0 | 0 | 0 | 1.3 | 1.2 | 0 | 0.1 | 0 | 0.5 | 517 |

## Forest areas

To represent the historical siting decisions of wind farms within carbon-rich ecosystems, we overlap the wind farm footprint dataset with geospatial data on forests. For this land-cover type, we employ the CORINE Land Cover dataset[43], specifically categories Broad-leaved forest (3.1.1), Coniferous forest (3.1.2), and Mixed forest (3.1.3). Table S5.4 presents the fractional overlap aggregated at the country level for forests.

Based on this information and the 15% threshold, we infer that, in most countries (excluding those with fewer than 5 wind farms), past siting decisions for wind projects have a varied importance, with countries like Austria, Croatia, Finland, Germany, Greece, Ireland, Italy, Sweden, and the United Kingdom siting wind farms on this land type. Therefore, we exclude all the forest areas in countries with less than 15% overlap.

**Table S5.4: Fractional overlap of wind farm footprint for forest land-cover type.**

| Country | Forest (%) | Num. of WFs |
|---|---|---|
| **Austria** | 16.2 | 43 |
| **Belgium** | 4.9 | 83 |
| **Bulgaria** | 3.1 | 49 |
| **Croatia** | 33.8 | 32 |
| **Czechia** | 13.3 | 48 |
| **Denmark** | 2.9 | 42 |
| **Estonia** | 55 | 3 |
| **Finland** | 80 | 141 |
| **France** | 4.1 | 506 |
| **Germany** | 22.6 | 490 |
| **Greece** | 16.8 | 87 |
| **Hungary** | 1.7 | 13 |
| **Ireland** | 22.2 | 52 |
| **Italy** | 20.5 | 38 |
| **Latvia** | 1 | 2 |
| **Lithuania** | 4.5 | 21 |
| **Luxembourg** | 23 | 4 |
| **Netherlands** | 2.1 | 80 |
| **Norway** | 12.4 | 52 |
| **Poland** | 3.5 | 170 |
| **Portugal** | 12.7 | 22 |
| **Romania** | 1.2 | 59 |
| **Slovakia** | 62.5 | 1 |
| **Slovenia** | 0 | 2 |
| **Spain** | 4.4 | 96 |
| **Sweden** | 80 | 119 |
| **Switzerland** | 5.1 | 18 |
| **United Kingdom** | 18.6 | 83 |

## Landscape visual impact considerations

To evaluate the aesthetic impact of wind energy deployment, we assess the visual impact of existing infrastructure by overlaying the wind farm footprints with high-resolution scenic landscape data across Europe[54,55]. This dataset comprises ternary classifications ("other", "scenic", and "highly scenic") of landscape scenicness predictions derived from machine learning models trained on the ScenicOrNot dataset from Great Britain. By intersecting these geospatial data with wind farm footprints, we quantify the extent to which onshore wind development has occurred in areas of high visual sensitivity or aesthetic value.

Table S5.5 shows the fractional overlap over each category across Europe. From this, we can see that most countries use "other" areas for wind energy deployment, followed by "scenic" areas, and only a few use "highly scenic" lands. Based on these values and the 15% threshold, we infer that most countries avoid siting wind farms in "highly scenic" areas, except in Greece, Ireland, Norway, Sweden, and the United Kingdom, whereas, for "scenic" areas, siting decisions are more varied. Countries such as Belgium, Czechia, Denmark, Finland, Greece, Ireland, Italy, Lithuania, Luxembourg, Portugal, Spain, Sweden, Switzerland, and the United Kingdom have consistently used scenic areas for wind energy development. For this category, we include all "other" areas for wind deployment, whereas "scenic" and "highly scenic" areas are excluded for countries with less than 15% overlap (excluding countries with fewer than 5 wind farms).

**Table S5.5: Fractional overlap of wind farm footprint aggregated by country for landscape visual impact.**

| Country | Other (%) | Scenic (%) | Highly scenic (%) | No classified data (%) | Num. of WFs |
|---|---|---|---|---|---|
| **Austria** | 93.9 | 6.1 | 0 | 0 | 85 |
| **Belgium** | 75.3 | 21.2 | 0.4 | 3.2 | 176 |
| **Bulgaria** | 95.5 | 4.5 | 0 | 0 | 49 |
| **Croatia** | 62.9 | 31 | 6.2 | 0 | 32 |
| **Czechia** | 74.5 | 25.4 | 0.1 | 0 | 48 |
| **Denmark** | 77.6 | 19.9 | 0.4 | 2.1 | 496 |
| **Estonia** | 73.1 | 26.9 | 0 | 0 | 3 |
| **Finland** | 66.1 | 28.9 | 2.8 | 2.2 | 141 |
| **France** | 86.7 | 12.3 | 1 | 0 | 1070 |
| **Germany** | 89 | 10.4 | 0.4 | 0.1 | 2383 |
| **Greece** | 16.6 | 47.7 | 33.3 | 2.5 | 142 |
| **Hungary** | 99.2 | 0 | 0 | 0.8 | 13 |
| **Ireland** | 49.8 | 31.3 | 17.7 | 1.1 | 162 |
| **Italy** | 71.6 | 22.9 | 5.5 | 0 | 259 |
| **Latvia** | 86.9 | 13.1 | 0 | 0 | 8 |
| **Lithuania** | 78.4 | 21.6 | 0 | 0 | 34 |
| **Luxembourg** | 67.3 | 32.7 | 0 | 0 | 10 |
| **Netherlands** | 96 | 2 | 0 | 2 | 178 |
| **Norway** | 1.9 | 8.4 | 86.4 | 3.3 | 64 |
| **Poland** | 90.8 | 9.1 | 0 | 0.1 | 320 |
| **Portugal** | 52.1 | 36.9 | 11 | 0 | 179 |
| **Romania** | 93.5 | 6.2 | 0.2 | 0.1 | 59 |
| **Slovakia** | 12.5 | 87.5 | 0 | 0 | 1 |
| **Slovenia** | 0.6 | 99.4 | 0 | 0 | 2 |
| **Spain** | 63.6 | 33.5 | 3 | 0 | 416 |
| **Sweden** | 44.2 | 38 | 16.6 | 1.3 | 721 |
| **Switzerland** | 52.4 | 47.4 | 0 | 0.2 | 18 |
| **United Kingdom** | 42.6 | 31.1 | 25.4 | 0.8 | 517 |

## Birds and bats vulnerability considerations

To assess how past practices have accounted for the impacts on bird and bat species in wind farm planning decisions, we combine data on the vulnerability of bird species to collisions with wind farms and on the distribution of threatened European bat species. For bats, we use a pseudo-species richness map for 10 threatened species (identified from Ref.[56]), obtained by stacking species distributions modelled in Ref.[57]. For birds, we use a novel dataset[58] that provides the number of species vulnerable to collisions at 10 km x 10 km resolution across Europe. Specifically, we use the pseudo-species richness map for 27 species classified as 'High risk' or 'High latent risk'. Further details on this dataset and the species considered are provided in Ref.[59]. We then combine the bird and bat pseudo-species richness maps and overlay the resulting layer with the wind farm footprint dataset.

In the medium scenario for this category, we first compute the percentile distribution for each country based on pseudo-species richness per grid cell. Next, for every 5% step in the distribution (up to the percentile 90th), cells with pseudo-species richness above this level (e.g. upper 10%, upper 15% and so on) are used to compute the fractional overlap with each country's wind farm footprints. This dataset then indicates at which point the number of vulnerable species within a country becomes relevant for wind farm siting, if at all. Table S5.6 shows the country-aggregated overlap between the wind farm footprints and the cells with pseudo-species richness above every percentile.

Countries such as Austria, Belgium, and Finland show no signal even for cells with the upper 10% (percentile 90th) pseudo-species richness, i.e., wind farms in these countries spatially overlap these cells more than the 15% threshold we apply. On the other hand, only France shows an overlap of 13% for grid cells that contain the upper 30% (percentile 70th) pseudo-species richness, potentially implying that it has planning processes which avoid these areas. As such, we use this dataset, apply the 15% overlap threshold to identify the percentile at which avoidance becomes significant and exclude grid cells with pseudo-species richness above this from development in the medium environmental scenario.

**Table S5.6: Fractional overlap of wind farm footprint aggregated by country for vulnerable bird and threatened bat pseudo-species richness.** The values in the table show the country-aggregated overlap between the wind farms and the cells with pseudo-species richness above every percentile.

| Country | 5th | 10th | 15th | 20th | 25th | 30th | 35th | 40th | 45th | 50th | 55th | 60th | 65th | 70th | 75th | 80th | 85th | 90th | Num. of WFs |
|---|---|---|---|---|---|---|---|---|---|---|---|---|---|---|---|---|---|---|---|
| **Austria** | 100 | 96.4 | 96.4 | 96.3 | 96.3 | 96.3 | 96.3 | 96.1 | 91 | 91 | 91 | 91 | 91 | 85.5 | 85.5 | 85.4 | 69.1 | 69.1 | 85 |
| **Belgium** | 99.5 | 99.5 | 99.5 | 99.5 | 84.5 | 84.5 | 84.5 | 67 | 67 | 65 | 58.6 | 54.4 | 48.5 | 48.5 | 36.3 | 36.3 | 25 | 25 | 175 |
| **Bulgaria** | 85.4 | 85.4 | 83.4 | 83.4 | 83.4 | 82.3 | 82.3 | 82.3 | 82.3 | 82.3 | 71.9 | 71.9 | 71.9 | 71.9 | 66.5 | 66.5 | 61.5 | 28.5 | 49 |
| **Croatia** | 98.6 | 98.6 | 98.6 | 98.6 | 98.6 | 98.6 | 98.6 | 98.6 | 98.6 | 98.6 | 98.6 | 98.6 | 85.7 | 85.7 | 85.7 | 85.7 | 72.2 | 72.2 | 32 |
| **Czechia** | 99.9 | 95.5 | 95.5 | 95.5 | 95.5 | 46.3 | 46.3 | 46.3 | 46.3 | 46.3 | 46.3 | 37.2 | 37.2 | 37 | 37 | 7.8 | 7.8 | 1.6 | 48 |
| **Denmark** | 100 | 100 | 83.5 | 82.2 | 82.2 | 78.1 | 78.1 | 71.6 | 71.6 | 62.2 | 62.2 | 62.2 | 50.9 | 50.9 | 50.9 | 50.9 | 20.8 | 20.8 | 496 |
| **Estonia** | 100 | 100 | 100 | 100 | 100 | 100 | 100 | 100 | 100 | 100 | 100 | 20.6 | 20.6 | 20.6 | 20.6 | 20.6 | 20.6 | 0 | 3 |
| **Finland** | 96.3 | 96.3 | 96.3 | 96.3 | 96.3 | 96.3 | 96.3 | 96.3 | 96.3 | 74.4 | 74.4 | 74.4 | 74.4 | 74.4 | 74.4 | 74.4 | 27 | 27 | 141 |
| **France** | 93.9 | 93.9 | 93.9 | 64.7 | 64.7 | 64.7 | 39.7 | 39.7 | 27.1 | 27.1 | 27.1 | 20.2 | 20.2 | 13.1 | 13.1 | 9.4 | 6.3 | 6.3 | 1061 |
| **Germany** | 97.9 | 92.7 | 92.7 | 92.7 | 77.9 | 77.9 | 77.9 | 62.1 | 62.1 | 62.1 | 62.1 | 44.4 | 44.4 | 44.4 | 44.4 | 30.6 | 30.6 | 21 | 2380 |
| **Greece** | 100 | 90.2 | 83.5 | 76.5 | 57.6 | 57.6 | 46.8 | 46.8 | 46.8 | 46.8 | 35.2 | 35.2 | 35.2 | 22.2 | 22.2 | 22.2 | 15.1 | 7.6 | 142 |
| **Hungary** | 100 | 100 | 100 | 96.1 | 96.1 | 96.1 | 96.1 | 78.4 | 78.4 | 78.4 | 78.4 | 39.6 | 39.6 | 39.6 | 39.6 | 22.8 | 22.8 | 22.8 | 13 |
| **Ireland** | 100 | 96.6 | 85.8 | 85.8 | 85.8 | 85.7 | 64.1 | 64.1 | 64.1 | 64.1 | 64.1 | 47.1 | 47.1 | 34 | 34 | 22.1 | 22.1 | 22.1 | 161 |
| **Italy** | 99 | 98.9 | 98.3 | 98.3 | 93.2 | 93.2 | 93.2 | 63.3 | 63.3 | 63.3 | 37.5 | 37.5 | 37.5 | 37.5 | 21.1 | 21.1 | 9.3 | 9.3 | 259 |
| **Latvia** | 100 | 100 | 100 | 100 | 100 | 100 | 100 | 100 | 100 | 100 | 100 | 100 | 100 | 100 | 34.4 | 34.4 | 34.4 | 34.4 | 8 |
| **Lithuania** | 99.9 | 99.9 | 99.9 | 99.9 | 99.9 | 99.9 | 86.7 | 86.7 | 86.7 | 86.7 | 86.7 | 86.7 | 86.7 | 67 | 67 | 67 | 16.3 | 16.3 | 34 |
| **Luxembourg** | 100 | 100 | 100 | 100 | 44.7 | 44.7 | 44.7 | 44.7 | 44.7 | 44.7 | 44.7 | 0 | 0 | 0 | 0 | 0 | 0 | 0 | 10 |
| **Netherlands** | 92.1 | 92.1 | 90.9 | 89.9 | 89.9 | 89.9 | 89.9 | 79.9 | 79.9 | 79.9 | 79.9 | 67.1 | 67 | 67 | 67 | 67 | 20.5 | 20.5 | 176 |
| **Norway** | 100 | 81.2 | 81.2 | 79.2 | 79.2 | 79.2 | 74.4 | 74.4 | 74.4 | 74.4 | 67.8 | 67.8 | 67.8 | 67.8 | 38.5 | 38.5 | 38.5 | 10.3 | 64 |
| **Poland** | 99.4 | 97.5 | 97.5 | 90.5 | 90.5 | 90.5 | 90.5 | 64.3 | 64.3 | 64.3 | 64.3 | 64.3 | 64.3 | 42.9 | 42.9 | 42.9 | 42.9 | 23.4 | 320 |
| **Portugal** | 94.2 | 94.2 | 76.4 | 60.1 | 60.1 | 60.1 | 60.1 | 30.1 | 30.1 | 30.1 | 30.1 | 9.9 | 9.9 | 9.9 | 9.9 | 2.6 | 2.6 | 2.6 | 177 |
| **Romania** | 99.8 | 99.8 | 99.8 | 99.8 | 99.8 | 99.8 | 99 | 99 | 99 | 99 | 98.9 | 98.9 | 98.9 | 98.9 | 96.2 | 90.4 | 89.9 | 71.6 | 59 |
| **Slovakia** | 100 | 100 | 100 | 100 | 100 | 100 | 100 | 100 | 100 | 100 | 100 | 0 | 0 | 0 | 0 | 0 | 0 | 0 | 1 |
| **Slovenia** | 100 | 100 | 100 | 0.6 | 0.6 | 0.6 | 0.6 | 0.6 | 0.6 | 0.6 | 0.6 | 0.6 | 0.6 | 0.6 | 0.6 | 0.6 | 0.6 | 0.6 | 2 |
| **Spain** | 95.1 | 91.3 | 91.3 | 80.7 | 80.7 | 70.6 | 70.6 | 70.6 | 56.4 | 56.4 | 56.4 | 36.8 | 36.8 | 36.8 | 22 | 22 | 22 | 10.2 | 414 |
| **Sweden** | 98 | 98 | 98 | 97 | 97 | 97 | 89.8 | 89.8 | 89.8 | 89.8 | 58.4 | 58.4 | 58.4 | 58.4 | 29.5 | 29.5 | 20.9 | 20.8 | 721 |
| **Switzerland** | 100 | 100 | 100 | 100 | 100 | 100 | 100 | 100 | 83.1 | 83.1 | 81.1 | 81.1 | 81.1 | 45.7 | 45.7 | 45.7 | 14.5 | 14.5 | 18 |
| **United Kingdom** | 100 | 93.5 | 83.8 | 83.8 | 83.8 | 67.4 | 67.4 | 67.4 | 67.4 | 47.6 | 47.6 | 47.6 | 36.4 | 36.4 | 36.4 | 25.2 | 25.2 | 15.5 | 515 |

## Summary of exclusions

**Table S5.7: Country-specific exclusions for each environmental and social factor in the medium tolerance scenarios.** Here excluded means onshore wind deployment is restricted in that country for the specific category while included means it would be permitted.

| **Country** | **Setback distances (m)** | **Natura 2000 (-)** | **CDDA (excluded category)** | **Forest (-)** | **Landscape Visual Impact (excluded category)** | **Birds and bats (excluded upper percentile of vulnerable species)** |
|---|---|---|---|---|---|---|
| **Austria** | 1063 | Excluded | All | Included | Scenic, highly scenic | No restriction |
| **Belgium** | 628 | Excluded | All | Excluded | Highly scenic | No restriction |
| **Bulgaria** | 500 | Included | All except "NotApplicable" | Excluded | Scenic, highly scenic | No restriction |
| **Croatia** | 700 | Included | All | Included | Highly scenic | No restriction |
| **Czechia** | 500 | Excluded | All | Excluded | Highly scenic | 20% |
| **Denmark** | 600 | Excluded | All | Excluded | Highly scenic | No restriction |
| **Estonia** | 628 | Excluded | All | Included | Highly scenic | No restriction |
| **Finland** | 950 | Excluded | All | Included | Highly scenic | No restriction |
| **France** | 500 | Excluded | All | Excluded | Scenic, highly scenic | 30% |
| **Germany** | 897 | Excluded | All | Included | Scenic, highly scenic | No restriction |
| **Greece** | 1000 | Included | All | Included | None excluded | 10% |
| **Hungary** | 700 | Excluded | All | Excluded | Scenic, highly scenic | No restriction |
| **Ireland** | 500 | Included | All | Included | None excluded | No restriction |
| **Italy** | 550 | Excluded | All | Included | Highly scenic | 15% |
| **Latvia** | 800 | Excluded | All | Included | Scenic, highly scenic | No restriction |
| **Lithuania** | 600 | Excluded | All | Excluded | Highly scenic | No restriction |
| **Luxembourg** | 628 | Excluded | All | Included | Highly scenic | No restriction |
| **Netherlands** | 400 | Excluded | All | Excluded | Scenic, highly scenic | No restriction |
| **Norway** | 700 | N/A | All | Excluded | None excluded | 10% |
| **Poland** | 700 | Excluded | All | Excluded | Scenic, highly scenic | No restriction |
| **Portugal** | 500 | Included | All | Excluded | Highly scenic | 40% |
| **Romania** | 450 | Excluded | All | Excluded | Scenic, highly scenic | No restriction |
| **Slovakia** | 1000 | Excluded | All | Included | Highly scenic | No restriction |
| **Slovenia** | 500 | Excluded | All | Included | Highly scenic | No restriction |
| **Spain** | 625 | Excluded | All | Excluded | Highly scenic | 10% |
| **Sweden** | 750 | Excluded | All | Included | None excluded | No restriction |
| **Switzerland** | 628 | N/A | All | Excluded | Highly scenic | 15% |
| **United Kingdom** | 628 | Excluded | All | Included | None excluded | No restriction |

# Supplemental references